\documentclass[10pt,letterpaper]{article}

\usepackage{graphicx}

\renewcommand{\baselinestretch}{2}

\usepackage[margin=2.5cm]{geometry}

\begin{document}

\setlength{\parindent}{0em}
\newcommand{\nind}{\setlength{\parindent}{0em}}
\newcommand{\ind}{\setlength{\parindent}{2em}}

{\sf \LARGE Large spin-orbit coupling in carbon nanotubes}

\vspace{0.5em}

G. A. Steele$^{1,*}$, F. Pei$^1$, E. A. Laird$^1$, J. M. Jol$^1$, H. B. Meerwaldt$^1$, L. P. Kouwenhoven$^1$

\vspace{1em}

{\em $^1$Kavli Institute of NanoScience, Delft University of
  Technology, PO Box 5046, 2600 GA, Delft, The Netherlands. 

$^*$Correspondence and requests for materials should be address to
  G.A.S. (email: g.a.steele@tudelft.nl).}

\vspace{1em}

{\bf 

It has recently been recognized that the strong spin-orbit interaction
present in solids can lead to new phenomena, such as materials with
non-trivial topological order. 
Although the atomic spin-orbit coupling
in carbon is weak, the spin-orbit
coupling in carbon nanotubes can be significant due to their curved surface.
Previous works have reported spin-orbit couplings in reasonable
agreement with theory, and this coupling strength has formed
the basis of a large number of theoretical proposals.
Here we report a
spin-orbit coupling in three carbon nanotube devices that is an order
of magnitude larger than measured before.  
We find a zero-field spin
splitting of up to 3.4 meV, corresponding to a built-in effective magnetic
field of 29 T aligned along the nanotube axis.
While the origin of the large spin-orbit coupling is not
explained by existing theories, its strength is promising for
applications of the spin-orbit interaction in carbon nanotubes
devices.}

\setlength{\parindent}{0em}
\vspace{3em}

In solids, spin-orbit coupling has recently become a very active
topic, in particular in the context of its role in a new class of
materials with a non-trivial topological
order\cite{kane2011topological,hasan2010colloquium,hsieh2008topological},
and its use to enable new control techniques in solid-state qubits
based on manipulating spins with electric
fields\cite{nowack2007coherent,nadj2010spin}. Due to the low atomic
number of the carbon nucleus, the spin-orbit interaction in carbon
materials is, in general, weak. An example of this is flat graphene,
in which intrinsic spin-orbit effects are expected to appear at energy
scales of only 1 $\mu$eV (10
mK)\cite{huertas2006spin,min2006intrinsic}. In carbon nanotubes,
however, the curvature of the surface breaks a symmetry that is
present in graphene. This broken symmetry enhances the intrinsic
spin-orbit coupling in carbon nanotubes compared to flat graphene,
with theoretical estimates predicting splittings on the order of
$100\ \mu$eV, an energy scale easily accessible in transport
measurements at dilution refrigerator temperatures, and recently
observed in
experiments\cite{kuemmeth2008coupling,churchill2009relaxation,jhang2010spin,jespersen2011so}.
Experiments so far have reported spin-orbit splittings typically in
the range of hundreds of $\mu$eV, and which were reasonably consistent
with theoretical predictions.

\setlength{\parindent}{2em}

Since its first experimental observation, the spin-orbit interaction
in carbon nanotubes has attracted significant theoretical attention,
and has been the basis of a large number of theoretical
proposals. Recent calculations predict that it enables fast electrical
spin manipulation in carbon nanotube spin qubits
\cite{bulaev2008spin,flensberg2010bends}, that it can couple to the
phase of Josephson supercurrents through Andreev bound states in
nanotube superconducting
junctions\cite{zazunov2009anomalous,lim2011josephson}, that it allows
the spin to couple to the high quality vibrational modes of
nanotubes\cite{palyi2012spin,ohm2012readout}, and that it could be
interesting for the study of topological liquids and Majorana bound
states
\cite{lutchyn2010majorana,oreg2010helical,klinovaja2011helical,egger2012emerging,sau2011majorana}. These
many exciting proposed applications could potentially benefit from a
stronger spin-orbit coupling.

Here, we present measurements of three carbon nanotube devices which
have spin-orbit couplings an order of magnitude larger than that
predicted by theory. We observe the spin-orbit coupling by measuring
the magnetic field dependence of the ground states of clean carbon
nanotube quantum dots in the few-electron and few-hole regime
\cite{steele2009tunable}. We use a Dirac-point crossing at a low
magnetic field as a tool for distinguishing orbital-type
coupling\cite{tsuneya2000spin,de2002spin,huertas2006spin} from the
recently predicted Zeeman-type
coupling\cite{izumida2009spin,jeong2009curvature,klinovaja2011carbon}. While
it is not understood why the spin-orbit coupling we observe is so much
larger than that predicted by tight-binding calculations, its large
magnitude is attractive for implementing the theoretical proposals for
using the carbon nanotube spin-orbit coupling for a wide range of new
experiments.

\setlength{\parindent}{0em}
\vspace{2em}
{\large \bf Results}
\vspace{1em}

{\bf Large spin orbit coupling in a few electron nanotube quantum
  dot.}  The devices are made using a fabrication technique in which
the nanotube is deposited in the last step of the fabrication.  Figure
1a shows a schematic of a single quantum dot device with three
gates. Figure 1b shows a scanning electron microscope (SEM) image of
device 1, taken after all measurements were completed. Similar to
previous reports\cite{steele2009tunable}, we are able to tune the
device to contain only a single electron (see Supplementary Note 1
and Supplementary Figures S1 and S2). An external magnetic field is
applied in-plane, perpendicular to the trench. As we do not control
the direction of the growth process, this magnetic field often has a
misalignment to the nanotube, but still contains a large component
parallel to the nanotube axis. All measurements were performed in a
dilution refrigerator with an electron temperature of 100 mK.

\setlength{\parindent}{2em}

In figures 1c--f, we show the magnetic field dependence of the Coulomb
peaks of the first four electrons in a carbon nanotube quantum dot in
device 1. In the few electron regime, we estimate the single-particle
level spacing of the quantum dot to be $\Delta E_{SP} = 11$ meV (see
Supplementary Figure S3). Note that similar to recent
reports\cite{deshpande2009mott}, this device exhibits a crossing of
the Dirac point at an anomalously low magnetic field, causing a
reversal of the orbital magnetic moment of one of the valleys at
$B_{Dirac} = 2.2$ T (see figures 2c--f).  The low $B_{Dirac}$
indicates a small shift of the $k_\perp$ quantization line from the
Dirac point (Figure 2a), and would predict a small electronic bandgap
contribution from the momentum $k_\perp$ of the electronic states
around the nanotube circumference: $E_{gap}^{k\perp} = 2 \hbar v_F
k_\perp = 7$ meV.  We describe a nanotube with a low Dirac-field
crossing as ``nearly metallic'', as the $k_\perp$ quantization line
nearly passes through the Dirac point. The bandgap in our device does
not vanish at $B_{Dirac}$, as would be expected, but instead retains a
large residual contribution $E_{gap}^{residual} = 80$ meV, similar to
previous reports\cite{deshpande2009mott}. It has been suggested that
this residual energy gap could arise from a Mott-insulating state,
although its exact origin remains a topic of investigation that we
will not address here. This low Dirac field crossing does not affect
the spin-orbit spectra we observe, and will later provide a unique
signature for distinguishing orbital
\cite{tsuneya2000spin,de2002spin,huertas2006spin} from Zeeman
\cite{izumida2009spin,jeong2009curvature,klinovaja2011carbon} type
coupling. We first focus on the behaviour at magnetic fields below
$B_{Dirac}$.

The unambiguous signature of the nanotube spin-orbit interaction can
be seen by comparing the low magnetic field behaviour in figures 1c
and d. Due to the opposite direction of circulation of the electronic
states about the nanotube circumference, the bandgap of the $K$ and
$K^\prime$ valleys both change in the presence of a parallel magnetic
field\cite{minot2004determination,jarillo2005electronic}. The bandgap
in one valley increases and the other decreases, both with a rate
given by $dE/dB = 2\mu_{orb}$, where $\mu_{orb} = d e v_{F\perp} / 4$
($\mu_{orb} \sim 220 \ \mu$eV / T for $d = 1$ nm). In the absence of
spin-orbit coupling, the first two electrons would both occupy the
valley with lower energy, and thus the first two ground states would
both shift down in energy with magnetic field. In figures 1c and d, we
observe a different behaviour: in particular, at low magnetic fields,
the second electron instead occupies the valley that is increasing in
energy with magnetic field. The occupation of the ``wrong'' valley by
the second electron is a result of the nanotube spin-orbit
interaction\cite{kuemmeth2008coupling}: The spin-orbit coupling in
nanotubes results in an effective magnetic field aligned along the
nanotube axis, which points in opposite directions for the $K$ and
$K^\prime$ valleys (Fig.\ 2d). This magnetic field produces a spin
splitting $\Delta_{SO}$ for the two spin species in the same
valley. In an external magnetic field, the second electron then enters
the ``wrong'' valley, and persists there until the energy penalty for
this exceeds $\Delta_{SO}$. In device 1, from the extract ground state
spectra shown in figure 3(a), we find a $\Delta_{SO} = 3.4 \pm 0.3$
meV. In addition to the ground state measurements, states consistent
with such a splitting have been observed in finite bias excited state
spectroscopy (see Supplementary Figures S3 and S4). We have also
observed a large $\Delta_{SO} = 1.5 \pm 0.2$ meV in a second similar
single-dot device (see Supplementary Figures S5-S9, and Supplementary
Note 2).

\vspace{1em}
\setlength{\parindent}{0em}

{\bf Spin-orbit coupling in nearly metallic carbon nanotubes.} In figure 2,
we show calculated energy levels of a nearly metallic carbon nanotube
including the spin-orbit interaction. In carbon nanotubes, there are
two contributions to the spin-orbit coupling, one which we describe as
orbital-type coupling, which induces a shift in the $k_\perp$
quantization
line\cite{izumida2009spin,jeong2009curvature,klinovaja2011carbon} and
results in an energy shift proportional to the orbital magnetic
moment. The second type, which we describe as Zeeman-type, shifts only
the energy of the electron spin with no shift in $k_\perp$. The energy
and momentum shifts from these couplings are illustrated in figures 2e
and f. Combining these two effects, we have the following Hamiltonian
for the spin-orbit interaction (equation 71 in
\cite{klinovaja2011carbon}):
\begin{equation}
H^{cv}_{SO} = \alpha S^z \sigma_1 + \tau \beta S^z
\end{equation}
where $S^z$ is the spin component along the axis of the nanotube,
$\sigma_1$ leads to a spin-dependent horizontal shift of the dispersion
relation along $k_\perp$ that is of opposite sign in different
valleys, while $\tau$ leads to a spin-dependent vertical shift that is
opposite in the two valleys.  The first term represents the
orbital-type of coupling, while the second represents the Zeeman-type
coupling. The coefficients $\alpha$ and $\beta$ determine the strength
of the two types of coupling, with $\Delta_{SO}^{orb} = \alpha =
(-0.08$ meV nm)$/r$ at $k_{||} = 0$, and $\Delta_{SO}^{Zeeman} = \beta
= (-0.31 \cos 3\theta$ meV nm)$/r$ where $\theta$ is the chiral angle
of the nanotube wrapping vector\cite{klinovaja2011carbon}, and $r$ is
the radius of the nanotube in nanometers. Through the $\cos(3\theta)$
term, $\Delta_{SO}^{Zeeman}$ is dependent on the chirality of the
nanotube, and is maximum for nanotubes with $\theta=0$, corresponding
to the zigzag wrapping vector. Direct experimental observation of the
Zeeman-type coupling has been, until now, difficult. There have been
two reported indications of a Zeeman-type coupling. The first is a
different $\Delta_{SO}$ for holes and
electrons\cite{izumida2009spin,jeong2009curvature}, which is not
present in the orbital-type spin-orbit
models\cite{tsuneya2000spin,de2002spin,huertas2006spin}. Such an
asymmetry was observed in the initial experiments by Kuemmeth {\em et
  al.}, and motivated in part the initial theoretical work predicting
the Zeeman-type coupling\cite{izumida2009spin,jeong2009curvature}. The
second indication is a scaling of $\Delta_{SO}$ over a large number of
electronic shells, as seen in recent
experiments\cite{jespersen2011so}, from which a small Zeeman-type
contribution was extracted.

\setlength{\parindent}{2em}

The low Dirac field crossing in the nearly-metallic carbon nanotubes
studied here provides a unique signature that allows us to identify
the type of coupling by looking at the energy spectrum of only a
single shell. In figure 2g, we show the calculated energy spectrum for
a nearly-metallic carbon nanotube with purely orbital-type coupling
(see Supplementary Note 3 for details of the model). Since the
orbital-type coupling shifts $k_\perp$, the spin-up and spin-down
states cross the Dirac point at significantly different magnetic
fields\cite{jhang2010spin}. For a purely Zeeman-type coupling, figure
2h, the two spin states cross the Dirac point at the same magnetic
field. By comparing the theoretical predictions in figures 2g and 2h
to the observed energy spectrum extracted from the Coulomb peaks in
figure 3a, we can clearly identify a Zeeman-type spin-orbit coupling,
suggesting that this nanotube has a chiral vector near $\theta =
0$. However, the magnitude of the spin-orbit splitting is much larger
than that predicted by theory (see Supplementary Table S1 and
Supplementary Note 4 for a summary of expected theoretical values and
previous experimental observations).  One possible origin for the
observed discrepancy is an underestimate of the bare atomic spin-orbit
coupling parameter from ab-initio calculations, which enters the
tight-binding calculations as an empirical input parameter.

In figure 3, we show the ground state energies of the first 12
electrons as a function of magnetic field, extracted from the Coulomb
peak positions (Supplementary Figure S9). The ground states energies
follow a four-fold periodic shell-filling pattern, with the spin-orbit
split energy spectrum reproduced in the second and third electronic
shell. In figure 3e, we plot the orbital magnetic moment as a function
of shell number, including a correction for the angle between the
magnetic field and the nanotube axis. As reported previously
\cite{jespersen2011gate}, the orbital magnetic moment changes with
shell number, an effect particularly strong in our device due to the
small $k_\perp$ implied by the low magnetic field Dirac crossing. In
figure 3f, we plot the observed $\Delta_{SO}$ as a function of the
orbital magnetic moment, together with the theoretical predictions
from equation 1.  In the plot, we have included the fact that the
orbital coupling coefficient $\alpha$ in equation 1 scales with the
orbital magnetic moment\cite{jespersen2011so}.  The green dashed line
shows the prediction from equation 1 for a nanotube with a 3 nm
diameter, emphasizing the disagreement between measured and the
theoretically predicted values. Also shown is the same prediction with
the coefficients scaled by a factor of 8 in order to obtain the order
of magnitude of the observed splitting.

Note that there are some discrepancies between the energy spectrum
extracted from the Coulomb peak positions (figure 3a-c) and the
theoretical spectra presented in figure 2.  The first discrepancy is a
small curvature of the extracted ground state energies at $B<0.15$ T
in figures 3a-c, which we attribute to artifacts from way in which the
magnetic field sweeps were performed (see Supplementary Note 5 and
Supplementary Figure S10).  The second discrepancy is a bending of the
extracted energies at $B<1.5$ T, particularly noticeable in the upper
two states of the second and third shells (blue and purple lines in
figures 3b,c), and a resulting suppressed slope for $B<1.5$ T in these
states.  Correlated with the gate voltages and magnetic fields where
the suppressed slopes occur, we observed a strong Kondo effect present
in the odd valleys (see Supplmentary Figure S2).  Due to the strong
tunnel coupling to the leads, the Kondo current in the valley can
persist up to fields of 1.5 T (see Supplementary Figure S9), and is
stronger in the higher shells where the tunnel coupling to the leads
is larger.  The model described in figure 2 does not include
higher-order effects, such as Kondo correlations, and it seems that it
is no able to correctly predict the position of the Coulomb peak in
these regions.  Qualitatively, the magnetic moments associated with
the states appear to be reduced by the strong Kondo effect, although
the reason for this is not understood.  Note that a suppressed
magnetic moment will reduce the apparent spin-orbit splitting, and
thus the large spin-orbit splittings reported here represent a lower
bound.

\vspace{1em}
\setlength{\parindent}{0em}

{\bf Large spin-orbit coupling in a nanotube double quantum dot.} In
figure 4, we present data from a third device in a p-n double quantum
dot configuration that also exhibits an unexpectedly large spin-orbit
coupling (see Supplementary Note 6 and Supplementary Figures S11 and
S12 for device details and characterization). Figures 4c and d show
measurements of the ground state energies of the first two electrons
and first two holes in the device as a function of parallel magnetic
field, measured by tracking the position of a fixed point on the bias
triangle in gate space (coloured circles in 4a) as a function of
magnetic field. The signature of the nanotube spin-orbit interaction
can be clearly seen by the opposite slope of the first two electrons
(holes) in Figure 4c (4d), and is consistent with the carbon nanotube
spin-orbit spectrum far from the Dirac crossing, shown in figure 4b.
The difference in the high magnetic field slopes corresponds to a
Zeeman splitting with $g \sim 2$, as expected from the spin-orbit
spectrum.  By calibrating the gate voltage shifts into energy using
the size and orientation of the finite bias triangles (see
Supplementary Note 6), we extract an orbital magnetic moment of
$\mu_{orb} = 0.8$ meV/T, a spin-orbit splitting $\Delta_{SO}^{1e} =
1.7 \pm 0.1$ meV for the first electron shell, and $\Delta_{SO}^{1h} =
1.3 \pm 0.1$ meV for the first hole shell. Estimating the diameter
from the orbital magnetic moment, theory would predict a
$\Delta_{SO}^{max} \sim 0.2$ meV for this device, an order of
magnitude below the observed values. Note that device 3 exhibits a
large spin-orbit coupling without a low $B_{Dirac}$, suggesting that
these two phenomena are not linked.

\setlength{\parindent}{2em}

From the slopes of the ground states, we predict that first two
electron levels will cross at a magnetic field $B_2 = \Delta_{SO} /
g\mu_B = 15$ T, while the first two hole levels do not cross.  The
crossing of the first two electron levels instead of the hole states,
as was observed by Kuemmeth {\em et al.}, implies the opposite sign of
the spin-orbit interaction, likely due to a different chirality of our
nanotube.  The absence of the low Dirac field crossing, however, does
not allow us to clearly separate the orbital and Zeeman contributions,
as was possible for the other two devices.

\setlength{\parindent}{0em}
\vspace{2em}
{\large \bf Discussion}
\vspace{1em}

We have observed strong spin-orbit couplings in carbon nanotubes that
are an order of magnitude larger than that predicted by theory, with
splittings up to $\Delta_{SO} = 3.4$ meV. By using a low Dirac field,
we are able to identify a strong Zeeman-type coupling in two
devices. The origin of the large magnitude of the spin-orbit splitting
observed remains an open question. Nonetheless, the observed strength
of the coupling is promising for many applications of the spin-orbit
interaction in carbon nanotube devices.

\setlength{\parindent}{0em}
\vspace{2em}
{\large \bf Methods}
\vspace{1em}

{\bf Sample Fabrication} The devices are made using a fabrication
technique in which the nanotube is deposited in the last step of the
fabrication. Single quantum dot devices were fabricated by growing the
device across predefined structures with three gates, using W/Pt
electrodes for electrical contacts to the nanotube, and a dry-etched
doped silicon layer to make gates\cite{steele2009tunable}. Double
quantum dot devices were fabricated by growing the nanotube on a
separate chip\cite{pei2012valley}. 

\vspace{1em}

{\bf Measurements} Measurements were performed with a base electron
temperature of 100 mK. For measurements performed with single quantum
dot devices, a magnetic field was applied with an orientation in the
plane of the sample, perpendicular to the trench. In measurements with
double quantum dot devices, a 3D vector magnet was used to align the
direction of the magnetic field along the axis of the nanotube. The
measurement datasets presented in this manuscript are available
online, see Supplemenatary Data 1.

\vspace{1em}

{\bf Extraction of the ground state energies} In order to convert
changes in gate voltage position of the Coulomb peak to changes in
energy of the ground state, a scaling factor $\alpha$ is required that
converts gate voltage shifts into an energy scale. This scaling factor
is measured by the lever-arm factor from the Coulomb diamond data,
such as that shown in Figure S3. In addition to the scaling of gate
voltage to energy, the ground state magnetic field dependence traces
must be offset by an appropriate amount, corresponding to subtracting
the Coulomb energy from the addition energy, to produce spectra such
as that shown in figure 3 of the main text. To determine this offset,
we use the fact that at $B=0$, time-reversal symmetry requires that
the electron states are two-fold degenerate. The offset for the 1e/2e
curves was thus chosen such that the extrapolated states are
degenerate at $B=0$. This was also used to determine the offset
between the 3e/4e curves. For the remaining offset between the 2e and
3e curves, we use the level crossing that occurs at $B_1$. At $B_1$,
the levels may exhibit a splitting due to intervalley scattering. This
results in a ground state energy which does not show a sharp kink at
$B_1$, but instead becomes rounded. The rounding of this kink in our
data, however, is small. We estimate $\Delta_{KK’} \sim 0.1$ meV, and
have offset the 2e/3e curves by this amount at the crossing at
$B_1$. The spin-orbit splittings are determined by the zero-field gap
in the resulting ground-state spectra. The error bars quoted on the
spin-orbit splittings are estimates based on the accuracy with which
the ground states energy curves can be aligned to produced plots such
as those in figure 3 of the main text.

\setlength{\parindent}{0em}


\setlength{\parindent}{0pt}

\vspace{1em} 

{\bf \large Acknowledgments}

We thank Daniel Loss, Jelena Klinovaja, and Karsten Flensberg for
helpful discussions. This work was supported by the Dutch Organization
for Fundamental Research on Matter (FOM), the Netherlands Organization
for Scientific Research (NWO), the EU FP7 STREP program (QNEMS).

\vspace{1em}

{\bf \large Author Contributions}

G.A.S., F.P., E.A.L., J.M.J., and H.B.M. performed the experiments;
G.A.S., F.P., and H.B.M. fabricated the samples; G.A.S. wrote the
manuscript; all authors discussed the results and contributed to the
manuscript.

\vspace{1em}

{\bf \large Additional Information}

Supplementary information accompanies this paper at
www.nature.com/naturecommunictions. Reprints and permission
information is available online at
http://npg.nature.com/reprintsandpermissions/.  Correspondence and
requests for materials should be addressed to G.A.S.

\vspace{1em}

The authors declare that they have no competing financial interests.

\pagebreak

\renewcommand{\baselinestretch}{1.0}

\begin{figure}
\begin{center}
\includegraphics[width=\textwidth]{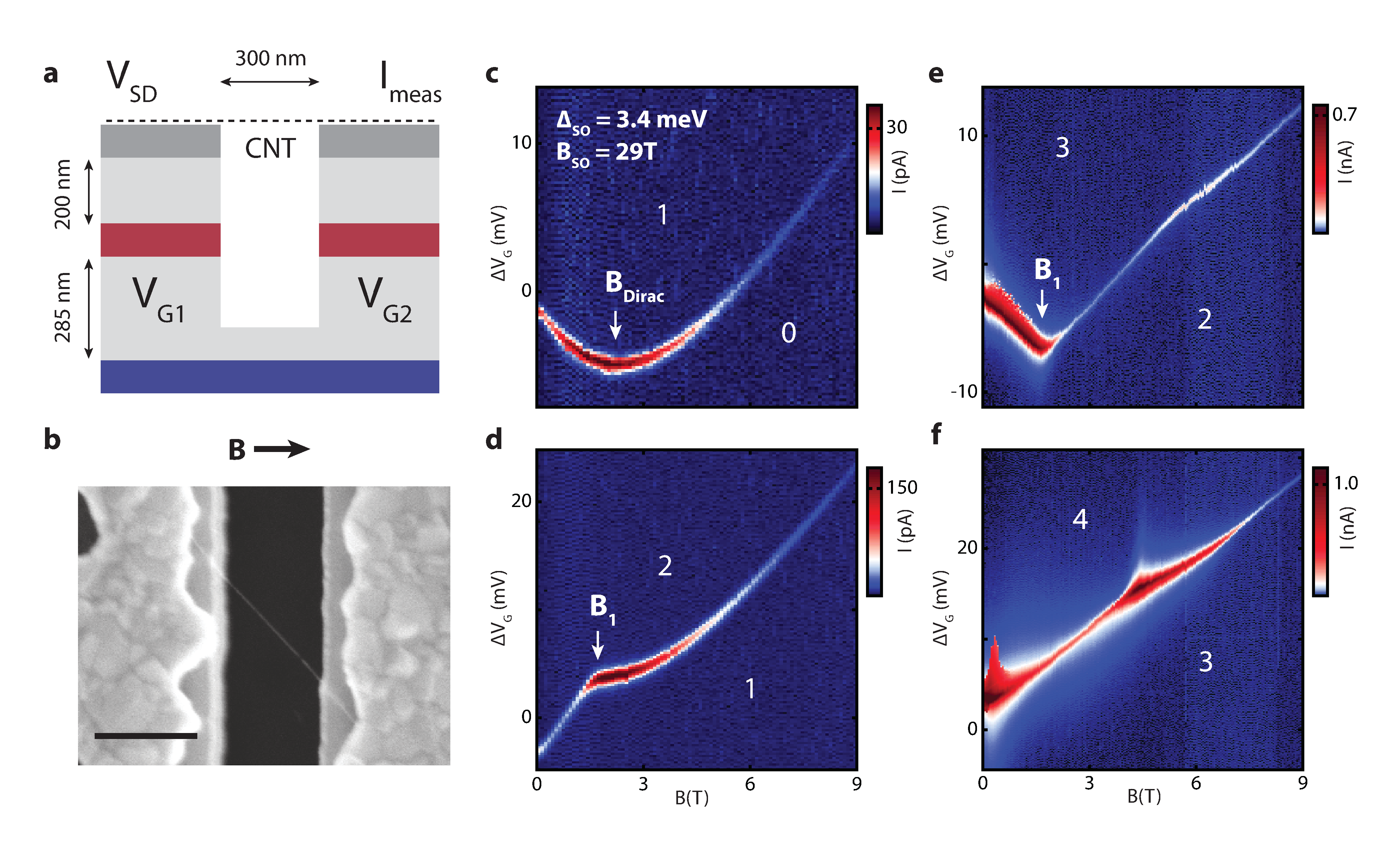}
\end{center}
\caption{ \textbf{A 29 T spin-orbit magnetic field in a carbon
    nanotube.} {\bf a,} A schematic of device 1. {\bf b,} A SEM image
  of device 1. Scale bar, 300 nm. Scale bar, 300 nm. The arrow
  indicates the direction of the applied magnetic field $B$. {\bf c}
  -– {\bf f,} Magnetic field dependence of the Coulomb peak positions
  of the first four electrons in the device. $V_{SD}$ = 200 $\mu$V in
  {\bf c,d} and $V_{SD}$ = 150 $\mu$V in {\bf e,f}. $\Delta V_G$
  corresponds to a small offset in gate voltage used to track the
  Coulomb peaks as a function of magnetic field. The crossing of the
  Dirac point reverses the sign of the orbital magnetic moment of the
  lower energy valley at a field $B_{Dirac} = 2.2$ T.  Without
  spin-orbit coupling, the first two electrons would both occupy the
  valley with the decreasing orbital energy, and would result in a
  downwards slope in both {\bf c} and {\bf d} at fields below
  $B_{Dirac}$. Here, the second electron, {\bf d}, instead occupies a
  valley with increasing orbital energy, a unique signature of the
  nanotube spin-orbit coupling, up to a field $B_1 = 1.6$ T. From the
  ground state energies extracted from the Coulomb peak positions,
  figure 3{\bf a}, we obtain a spin-orbit splitting $\Delta_{SO} = 3.4
  \pm 0.3$ meV, corresponding to a built-in spin-orbit magnetic field
  $B_{SO} = 29$ T seen by the electron spin. The sharp kinks at $B_1$
  in {\bf d} and {\bf e} imply weak valley mixing: we estimate
  $\Delta_{KK'} \sim 0.1$ meV.}
\end{figure}

\begin{figure}
\begin{center}
\includegraphics[width=\textwidth]{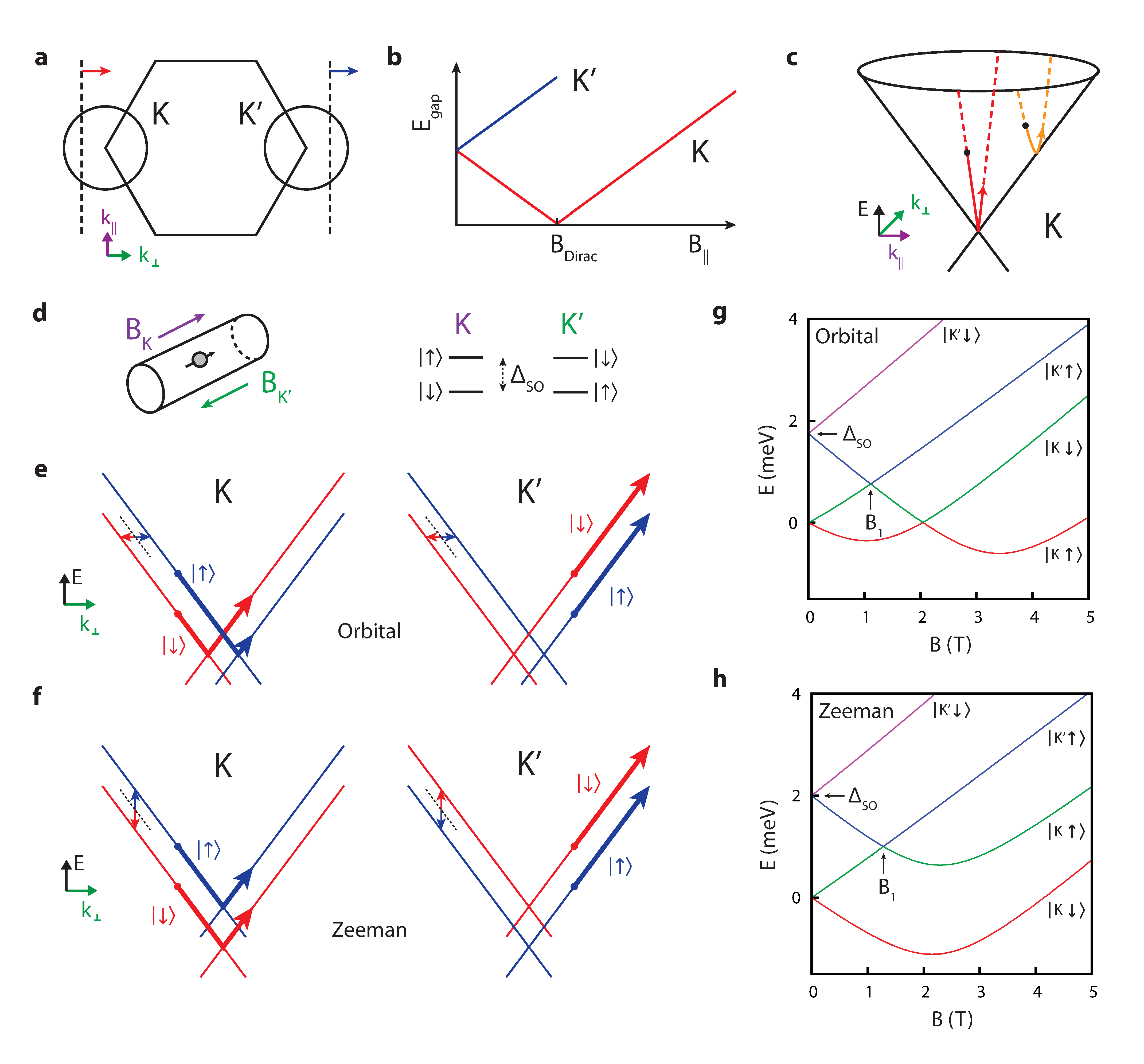}
\end{center}
\caption{ \textbf{Spin-orbit coupled states in a nearly metallic
    nanotube.} {\bf a} The two nanotube valleys (K and K$^\prime$)
  arise from the intersection of the $k_\perp$ quantization lines
  (dashed) with the Dirac cones of the graphene bandstructure. A
  magnetic field applied parallel to the nanotube axis shifts both
  quantization lines horizontally, reducing the bandgap in one K point
  and increasing it for the other, illustrated in {\bf b}. At a
  sufficiently large magnetic field $B_{Dirac}$, one valley (red line)
  crosses the Dirac point, after which the orbital magnetic moment
  changes sign. {\bf c,} With $k_\parallel$ = 0, the lowest energy
  state in the conduction band would follow a v-shape with a sharp
  kink at $B_{Dirac}$ (red line in {\bf b} and {\bf c}). A finite
  $k_\parallel$ from confinement in the axial direction results
  instead in a hyperbolic shape (orange line in {\bf c}). {\bf d,} The
  spin-orbit interaction in the nanotube results in an internal
  magnetic field aligned along the nanotube axis whose direction
  depends on the valley the electron occupies. {\bf e,} In the
  orbital-type spin-orbit
  coupling\cite{tsuneya2000spin,de2002spin,huertas2006spin}, this
  magnetic field results in a spin-dependent shift of $k_\perp$, while
  the Zeeman-type coupling, {\bf f}, gives a valley dependent vertical
  shift in
  energy\cite{izumida2009spin,jeong2009curvature,klinovaja2011carbon}. {\bf
    g, h,} Calculated energy spectrum of the first shell for a purely
  orbital-type coupling, {\bf g}, and a purely Zeeman-type coupling,
  {\bf h}, with parameters chosen to illustrate the difference between
  the two types of spectra. Colours indicate the ground state energies
  of the four electrons that would fill the shell. In {\bf g},
  electrons experience a spin-dependent $k_\perp$ shift, resulting in
  two separate Dirac crossings\cite{jhang2010spin}, an effect absent
  in {\bf h}.}

\end{figure}

\begin{figure}
\begin{center}
\includegraphics[width=\textwidth]{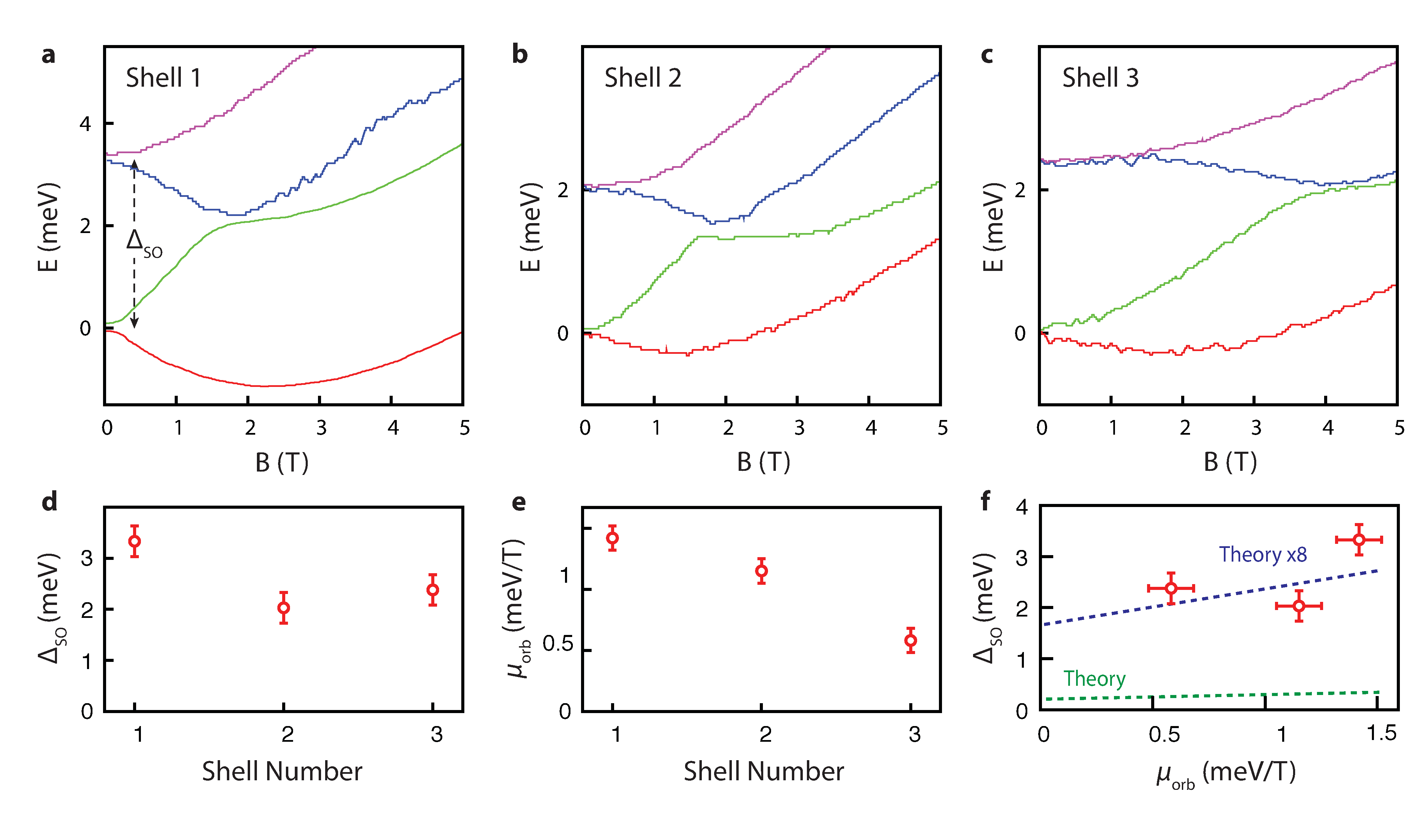}
\end{center}
\caption{\textbf{Spin-orbit coupling in the first three electronic
    shells.} {\bf a}-{\bf c}, Observed energy spectra of the first
  twelve electrons in device 1. The spectra exhibits a four-fold shell
  filling, with the spin-orbit electronic spectrum visible in all
  three shells. Extracted $\Delta_{SO}$ are shown in {\bf
    d}. Comparing to the spectra for the two types of nanotube
  spin-orbit coupling (fig. 2{\bf g} and 2{\bf h}), it is clear the
  device exhibits a Zeeman-type coupling. Deviations from the model
  are discussed in the main text. {\bf e,} $\mu_{orb}$ as a function
  of the shell number. For larger shells, electrons are confined in an
  electronic level with a larger value of $k_\parallel$. The
  correspondingly larger momentum along the nanotube axis decreases
  the velocity around the nanotube circumference, reducing the orbital
  magnetic moment\cite{jespersen2011so,jespersen2011gate}. {\bf f,}
  $\Delta_{SO}$ as a function of $\mu_{orb}$. The green dashed line
  shows the maximum spin-orbit coupling expected from theory with
  $\alpha$ and $\beta$ for a 3 nm nanotube (equation 1 together with
  the scaling of $\alpha$ with the magnetic moment). By scaling
  coefficients $\alpha$ and $\beta$ by a factor of 8 (blue line), we
  can reproduce the order of magnitude of the spin-orbit coupling in
  our device.}
\end{figure}

\begin{figure}
\begin{center}
\includegraphics[width=\textwidth]{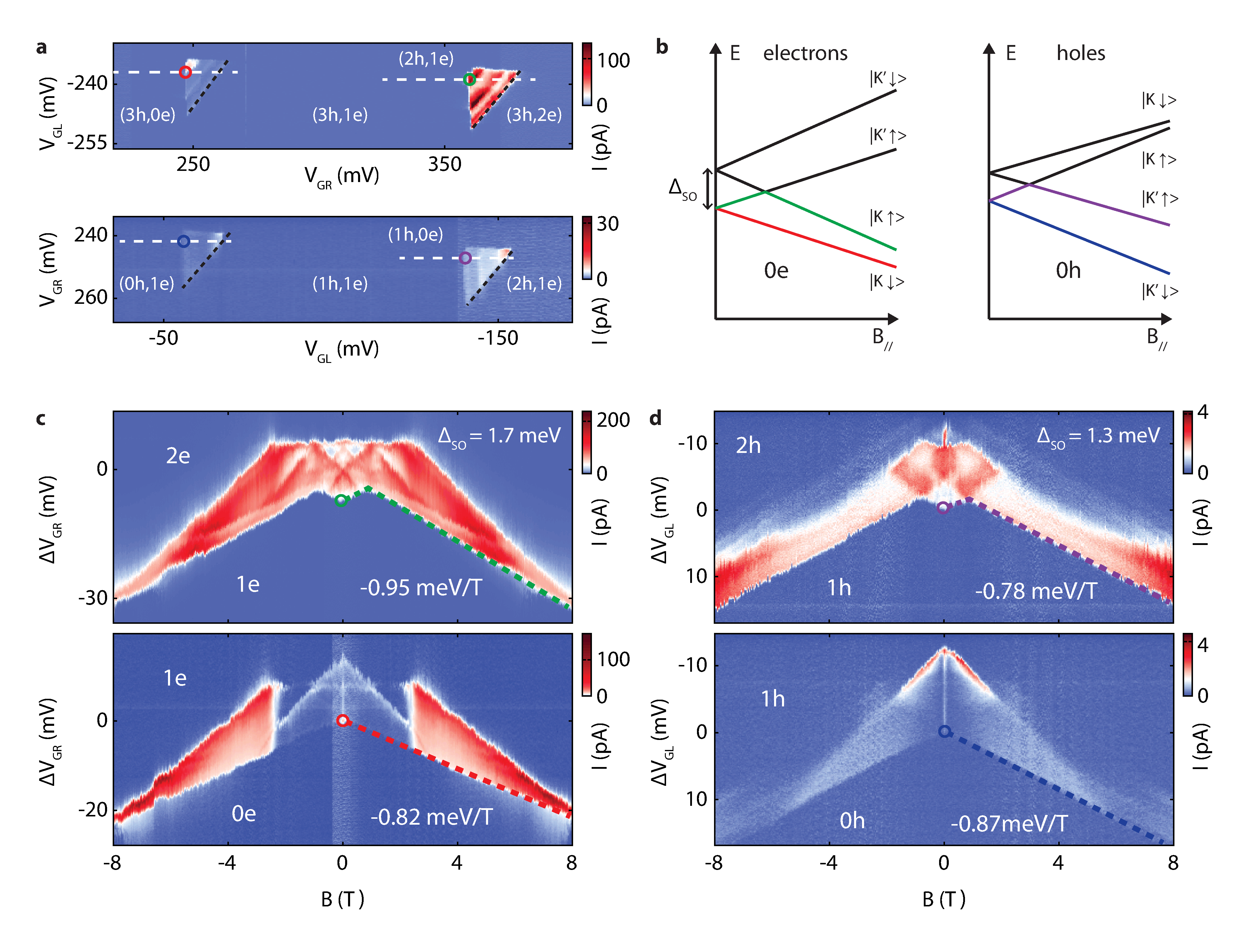}
\end{center}
\caption{{\bf Large spin-orbit coupling in a (p,n) double quantum
    dot.}  {\bf a,} Colourscale plots of the current at a source-drain
  bias $V_{sd}$ = 5 mV and $B=0$. Black dashed lines indicate the
  baseline of the triple-point bias triangles. Movement of the tip of
  the bias triangles (coloured circles) in gate space along line cuts
  in gate space (white dashed lines) with magnetic field is used in
  {\bf c} and {\bf d} to track the ground state energies. {\bf b,}
  Expected energy spectrum for the first shell of electrons and holes
  including the spin-orbit interaction. {\bf c,d,} Magnetic field
  dependence of line cuts in gate space (white dashed lines in {\bf
    a}) for the first two electrons, {\bf c}, and holes, {\bf
    d}. Coloured circles indicate positions on the corresponding bias
  triangles in {\bf a}, and the dashed lines indicate the observed
  magnetic field dependence of the ground states, in good agreement
  with the spin-orbit spectrum, {\bf b}. High magnetic field slopes
  for the ground state energies are indicated in the figures.  We
  extract spin-orbit splittings $\Delta_{SO}^{1e} = 1.7 \pm 0.1$ meV
  for the first electron shell and $\Delta_{SO}^{1h} = 1.3 \pm 0.1$
  meV for the first hole shell. Excited states inside the bias
  triangles (colourscale data above dashed lines) exhibit a rich
  structure as a function magnetic field, which we discuss elsewhere
  \cite{pei2012valley}.}
\end{figure}

\end{document}


\vspace*{\fill}

\begin{center}
\includegraphics[width=\textwidth]{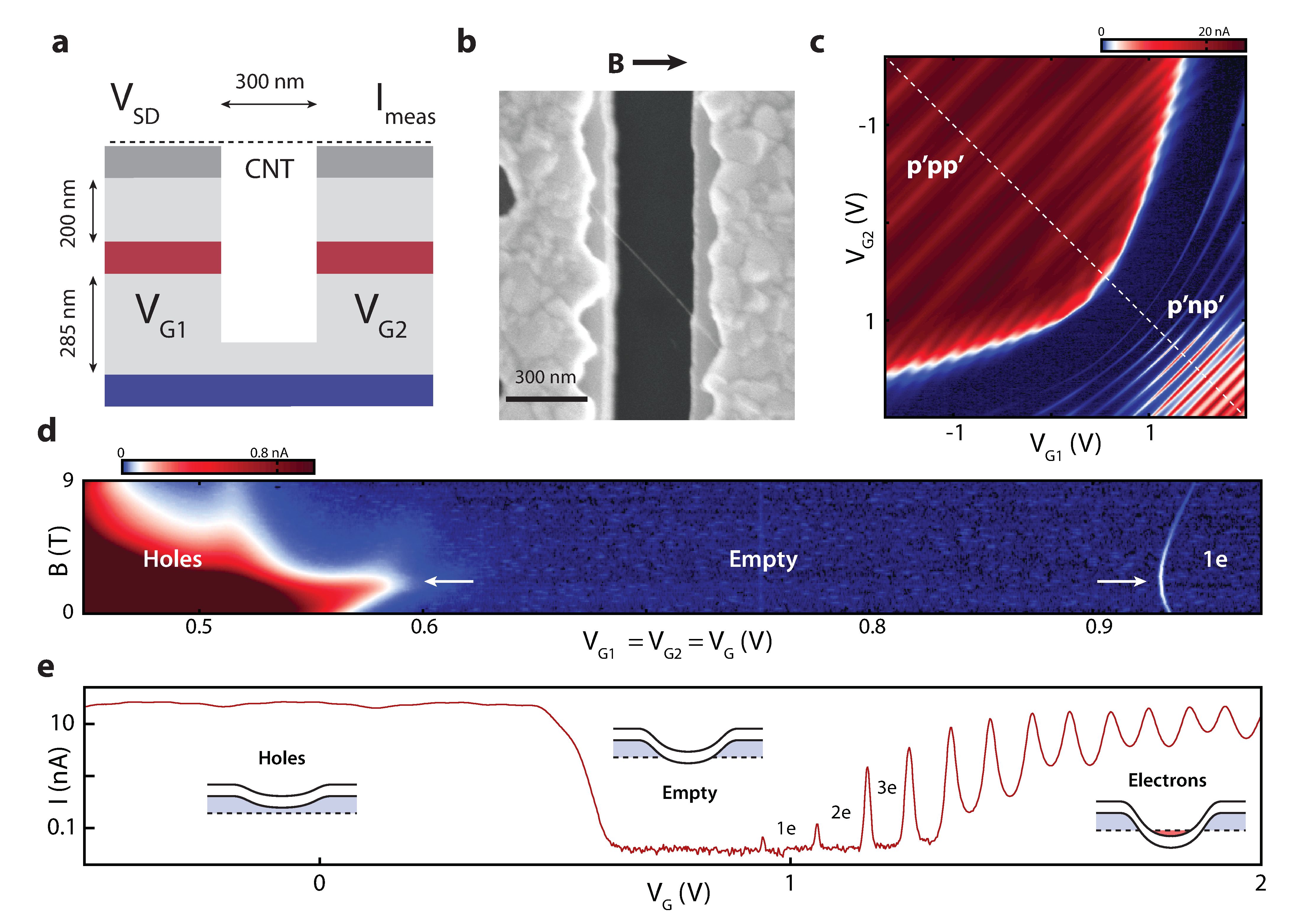}
\end{center}

\nind

{\bf Supplementary Figure S1: Characterization of Device 1.} {\bf a,}
Schematic of device 1, consisting of a clean nanotube grown over a
predefined trench. Source and drain contacts to the nanotube are made
by a 5/25 nm W/Pt bilayer (dark grey). Two gates embedded in the oxide
(red) are used to induce charges in the nanotube. A backgate (blue) is
kept grounded. {\bf b,} A scanning electron microscope (SEM) image of
the actual device, taken after all measurements. The nanotube axis
lies at an angle of 48 degrees relative to the magnetic field
orientation. {\bf c,} $I_{meas}$ at a $V_{sd} = 1$ mV as a function of
the two gate voltages. {\bf d,} Magnetic field dependence of
$I_{meas}$ along a diagonal line cut $V_{g1} = V_{g2} = V_g$ in {\bf
  c}. The device exhibits a minimum gap at $B_{Dirac} = 2.2$ T (white
arrows), corresponding to a crossing through the Dirac point of the
graphene bandstructure by the $k_\perp$ quantization line. {\bf e,}
Plot of $I_{meas}$ (logscale) at $V_{sd}$ = 1 mV showing the gate
voltages corresponding to the first electrons in the quantum dot.

\vspace*{\fill} \newpage \vspace*{\fill}

\begin{center}
\includegraphics[width=\textwidth]{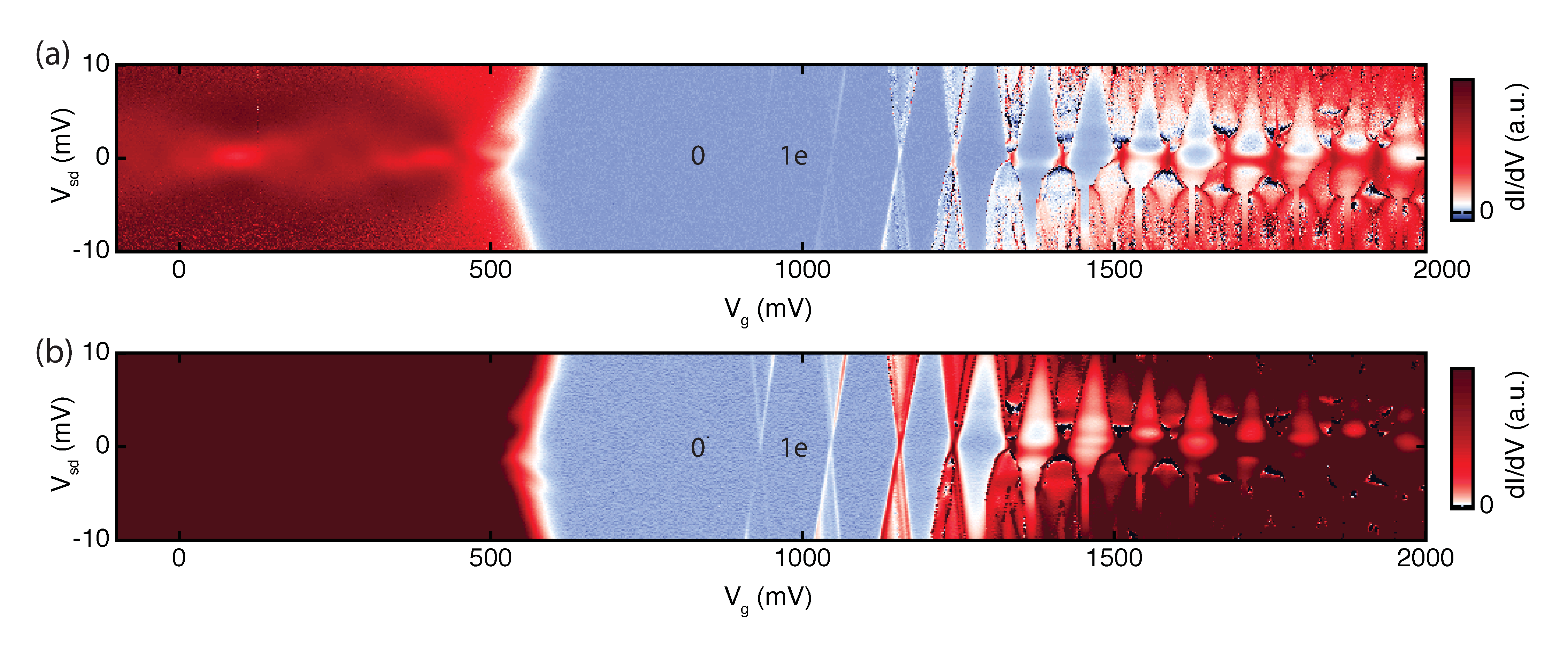}
\end{center}

{\bf Supplementary Figure S2: Stability diagram showing Coulomb
  diamonds of Device 1.} {\bf a,} Differential conductance of device 1
showing the Coulomb diamonds of the first electrons, the empty device,
and the threshold for hole conduction, taken at $B = 0$. For larger
gate voltage, a four-fold pattern of Kondo resonances is observed,
together with strong instabilities in the Coulomb diamonds which we
attribute to mechanical excitation of mechanical resonances of the
suspended nanotube by single-electron tunnelling [33]. Due to the lack
of p-n junction barriers for holes, the hole doped region shows
Fabry-Perot type oscillations. For the hole doped device, and for
large electron numbers, we estimate the single-particle energy of the
confined states to be $\sim 5$ meV. {\bf b,} The same data as in {\bf
  a} with the contrast enhanced in order to clearly show the Coulomb
diamond of the first electron.

\vspace*{\fill} \newpage \vspace*{\fill}

\begin{center}
\includegraphics[width=0.6\textwidth]{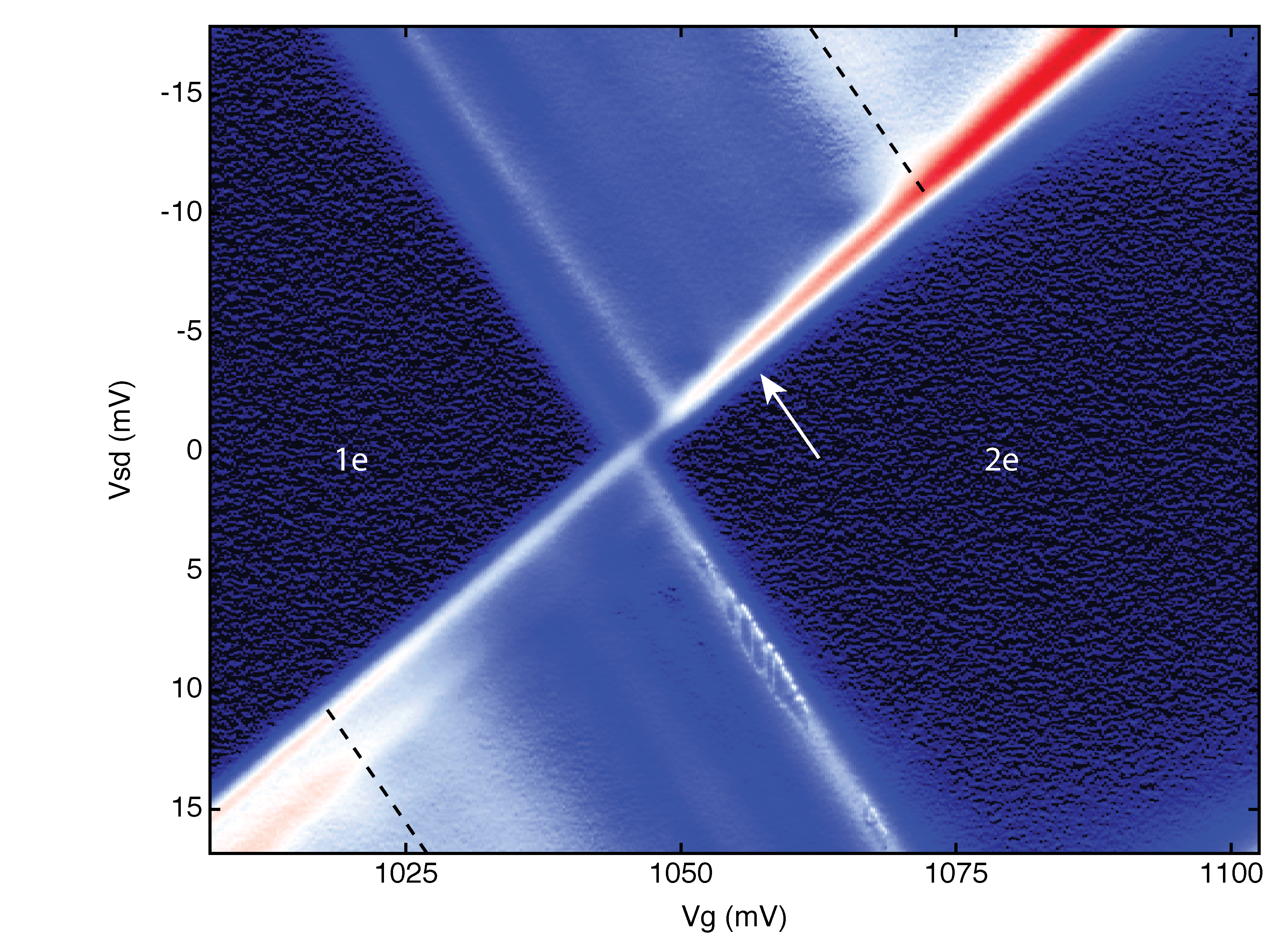}
\end{center}

{\bf Supplementary Figure S3: Excited states of the 2e charge state in
  Device 1} Differential conductance vs. $V_G$ and $V_{SD}$ for the 1e
to 2e transition, in which excited states could be resolved. The
dashed lines indicates excited states we identify as the
single-particle energy splitting $\Delta E_{SP}$. For the 1e-2e
transition, we extract $\Delta E_{SP} = 11$ meV. From $Delta E = \hbar
v_F / (2L)$, we estimate the size of the quantum dot $L \sim 200$
nm. From the measured angle in the SEM image, the total length of the
nanotube over the trench is $\sim 400$ nm. This implies a 100 nm
length for the pn depletion region and p’ doped regions from the work
function induced doping for the 1e quantum dot. For higher electron
numbers, and similarly for holes, the single particle energy drops to
$~ 5$ meV (see figure S4), implying a confinement length responding to
the full length of the suspended nanotube.  The white arrow indicates
the position of a faint excited state with an energy $\sim 3$ meV,
consistent with the spin-orbit splitting we observe from the ground
state measurements.

\vspace*{\fill} \newpage \vspace*{\fill}

\begin{center}
\includegraphics[width=0.8\textwidth]{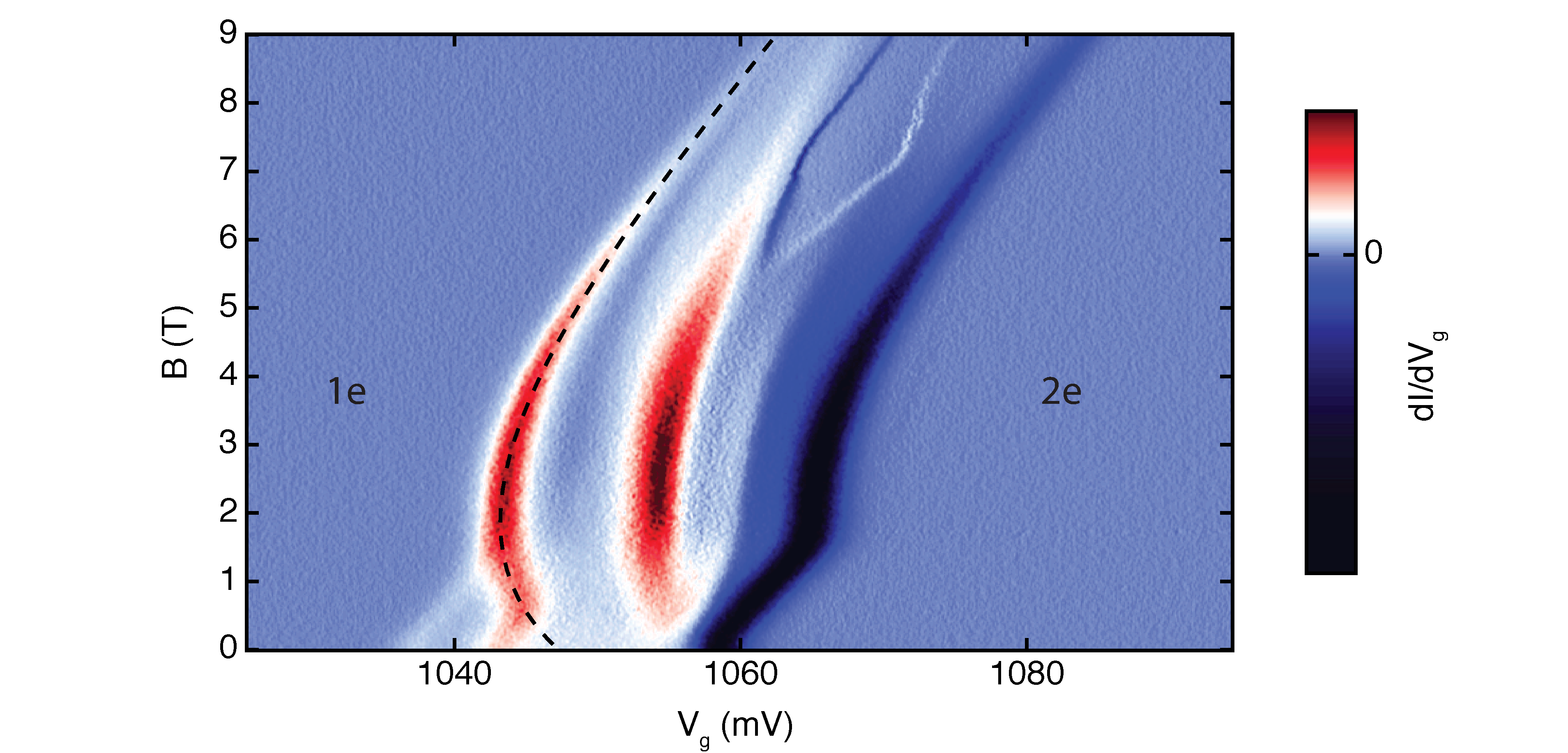}
\end{center}

{\bf Supplementary Figure S4: Evidence for spin-orbit splitting in
  magnetic field spectroscopy of excited states in Device 1.} A
colourscale plot showing $dI/dV_g$ as a function of magnetic field for
the 1e/2e transition of device 1 taken at $V_{SD} = 5.5$ mV. Excited
states of the 2e ground state appear as positive peaks in
$dI/dV_g$. The dashed line indicates the magnetic field dependence of
a 2e excited state consistent with the expected spectrum from
spin-orbit splitting.  From the excited state data, we extra a
spin-orbit splitting $\Delta_{SO} = 2.9$ meV, lower than that from the
ground state measurements. This difference can arise from a difference
between the excited state energies of the 2e state compared to the
ground state energy of 3e from electron interactions.  The origin of
the extra excited state running parallel to the ground state for
fields less than $B_1$ is not understood.  The value of $\Delta_{SO}$
from the excited state measurement is, similar to that from the ground
state measurements, an magnitude larger than the expected maximum
$\Delta_{SO}^{max} = 106$ $\mu$eV expected from theory (see table
S1). Excited states in the 0/1e transition did not show clear
visibility, and those of higher states were masked by instabilities we
attribute to mechanical excitation of the suspended nanotube (also
visible here above 6T).

\vspace*{\fill} \newpage \vspace*{\fill}

\begin{center}
\includegraphics[width=\textwidth]{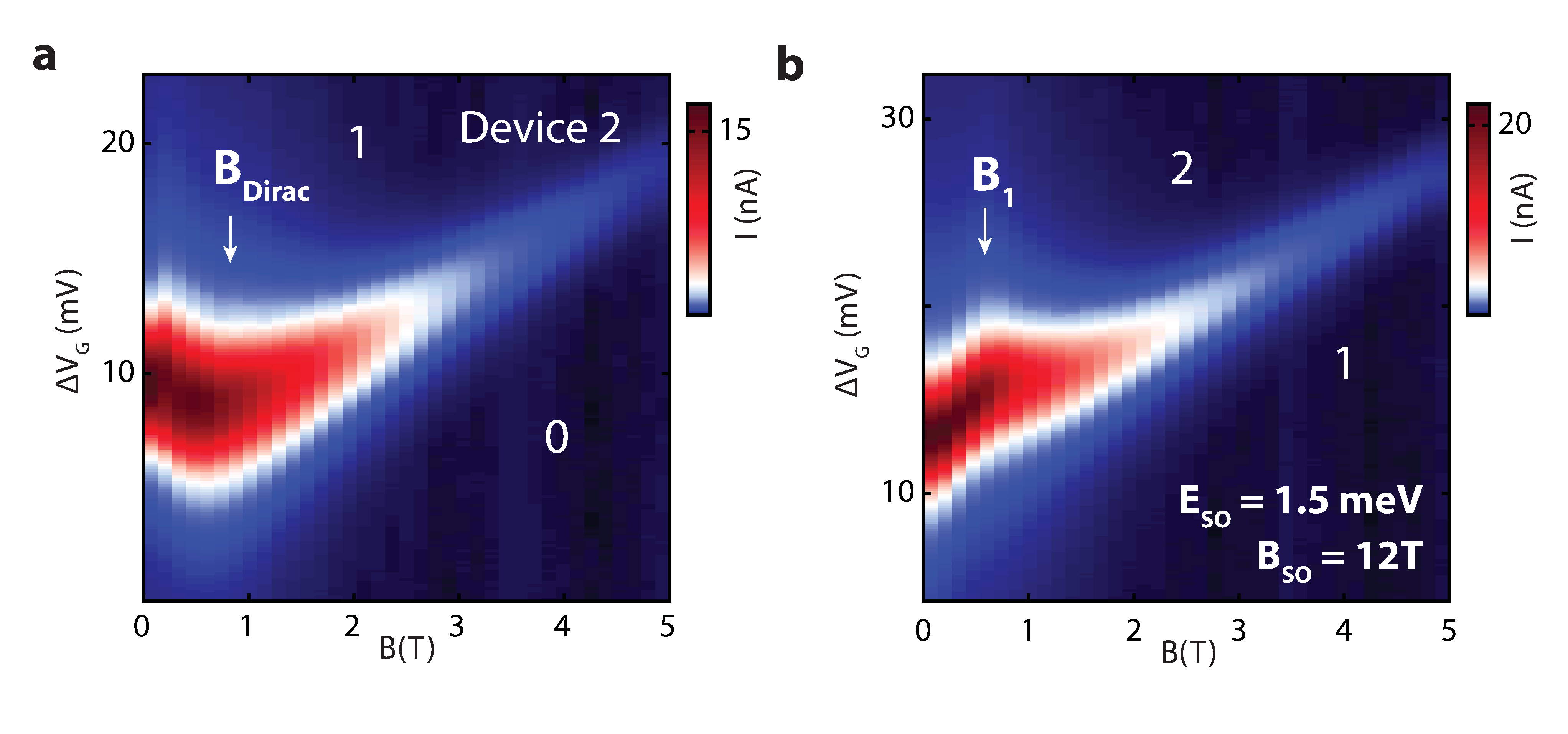}
\end{center}

{\bf Supplementary Figure S5: Spin-orbit split states in Device 2.}
Magnetic field dependence of the first two electrons in device 2,
showing the signature of the nanotube spin-orbit coupling with
$\Delta_{SO} = 1.5$ meV.

\vspace*{\fill} \newpage \vspace*{\fill}

\begin{center}
\includegraphics{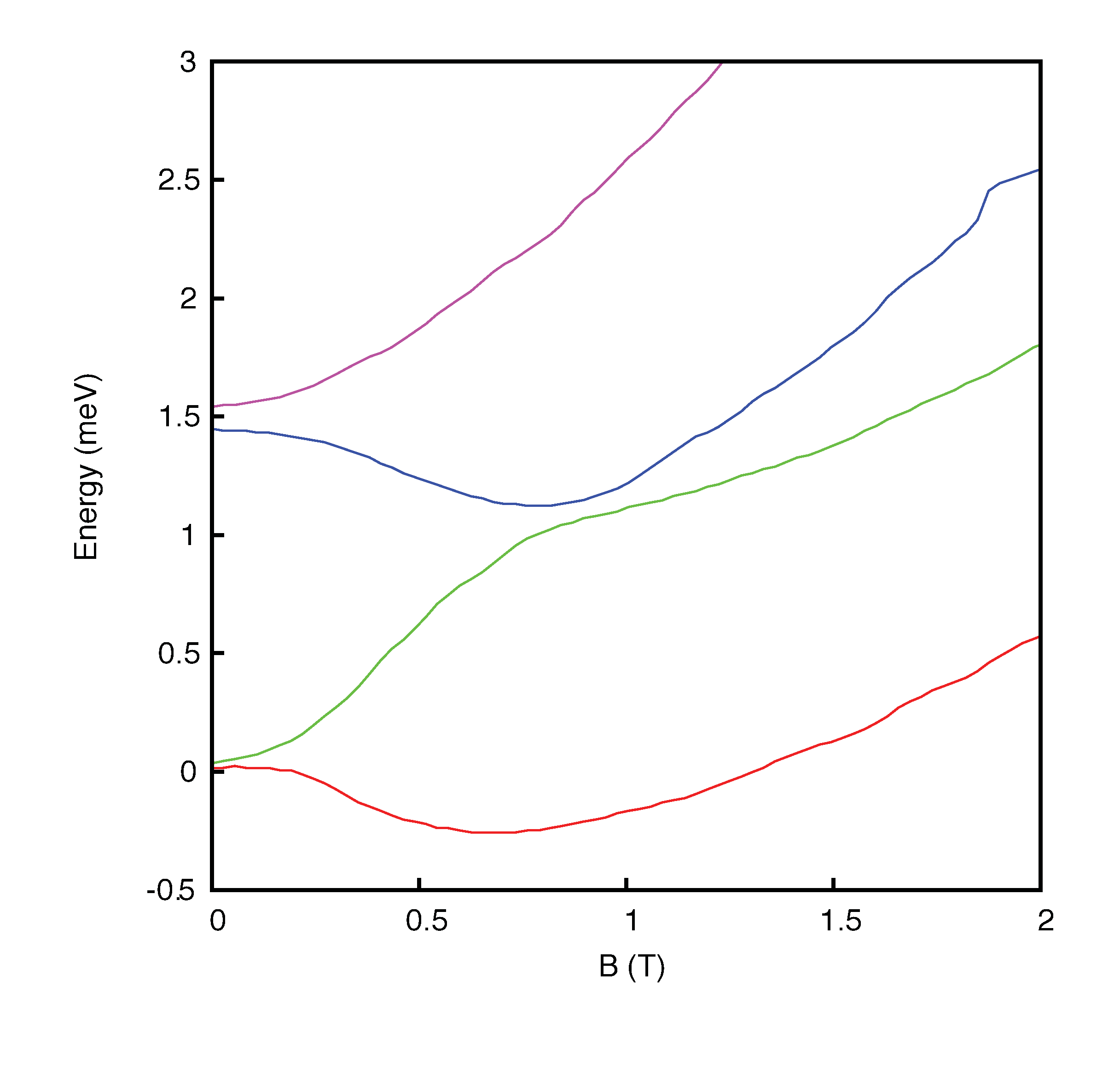}
\end{center}

{\bf Supplementary Figure S6: Spin-orbit spectrum of the first shell
  in Device 2.} Extracted ground state energies of the first four
electrons in device 2. Device 2 also shows a large spin orbit coupling
with a dominant Zeeman-type contribution. Note that the flat behaviour
of the ground states at zero magnetic field is a measurement artifact
from the magnetic field controller, see text for discussion.

\vspace*{\fill} \newpage \vspace*{\fill}

\begin{center}
\includegraphics[width=\textwidth]{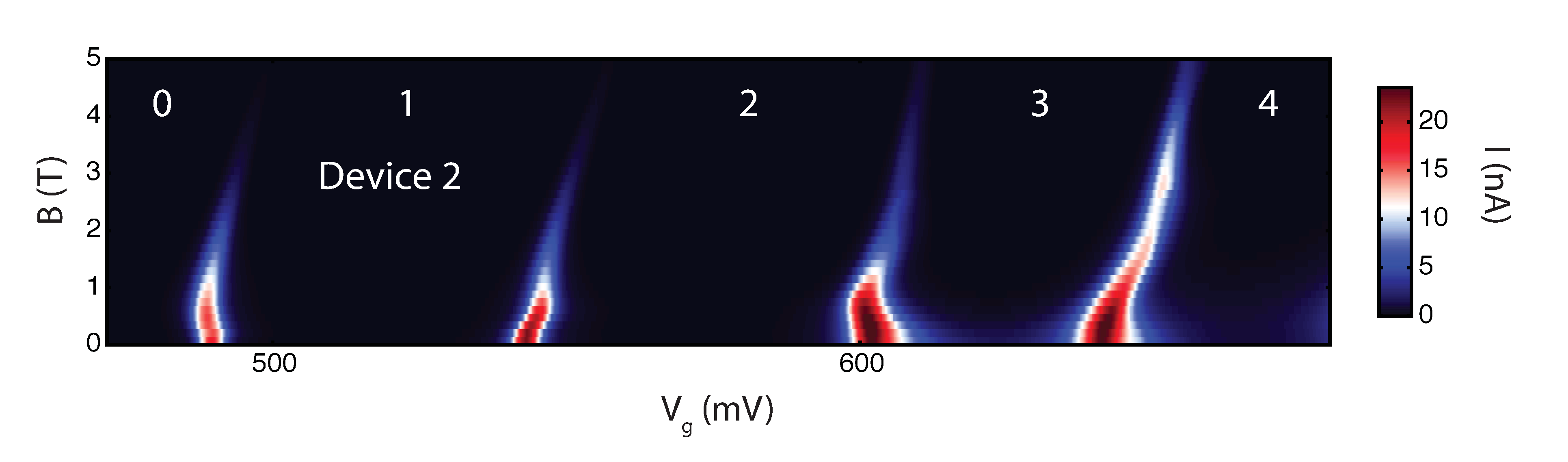}
\end{center}
{\bf Supplementary Figure S7: Magnetic field dependence of the first
  four Coulomb peaks in Device 2.} Coulomb blockade current
vs. magnetic field and gate voltage for device 2, used to extract the
ground state energies of the first shell, $V_{sd} = 1$ mV. The data in
figure S6 is a zoom of the data here.

\vspace*{\fill} \newpage \vspace*{\fill}

\begin{center}
\includegraphics[width=\textwidth]{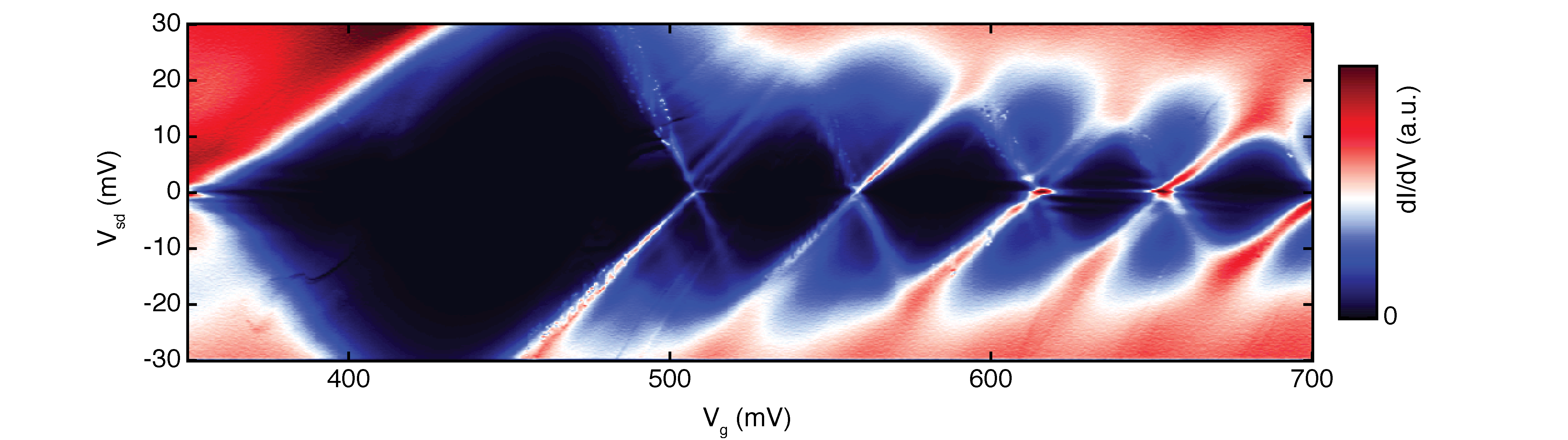}
\end{center}
{\bf Supplementary Figure S8: Stability diagram of Device 2 in
  few-electron regime.} {\bf a,} Differential conductance of device 2
showing the Coulomb diamonds of the first electrons, the empty device,
and the threshold for hole conduction, taken at $B = 0$.

\vspace*{\fill} \newpage \vspace*{\fill}

\includegraphics[width=\textwidth]{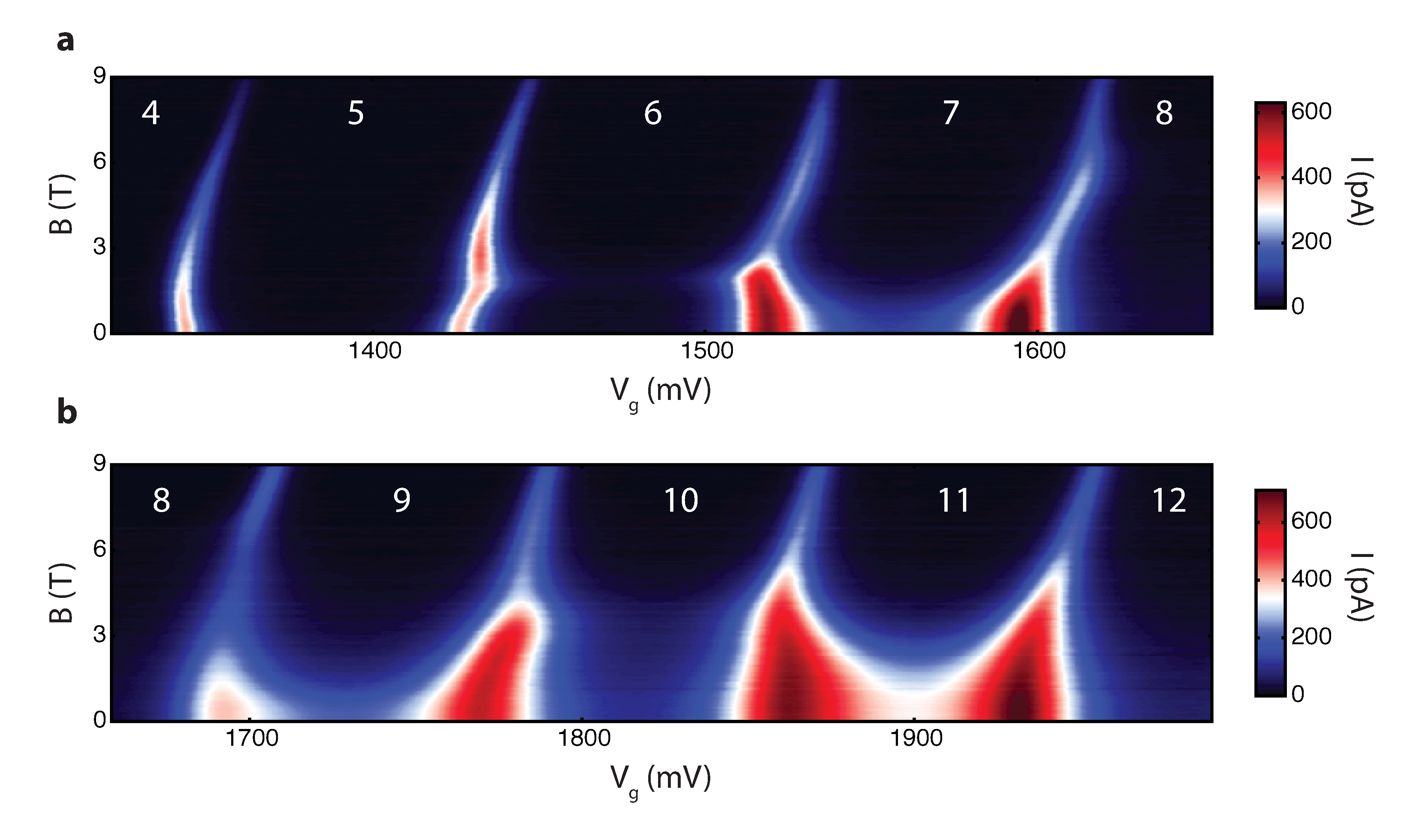} 

{\bf Supplementary Figure S9: Coulomb peak data for Shells 2 and 3 of
  Device 1.} Measurements of $I_{meas}$ vs. $Vg$ and $B$ on device 1
that are used to extract the energy spectra shown in figure 3 of the
main text, taken at $V_{sd}$ = 70 $\mu$V.

\vspace*{\fill} \newpage \vspace*{\fill}

\begin{center}
\includegraphics[height=6in]{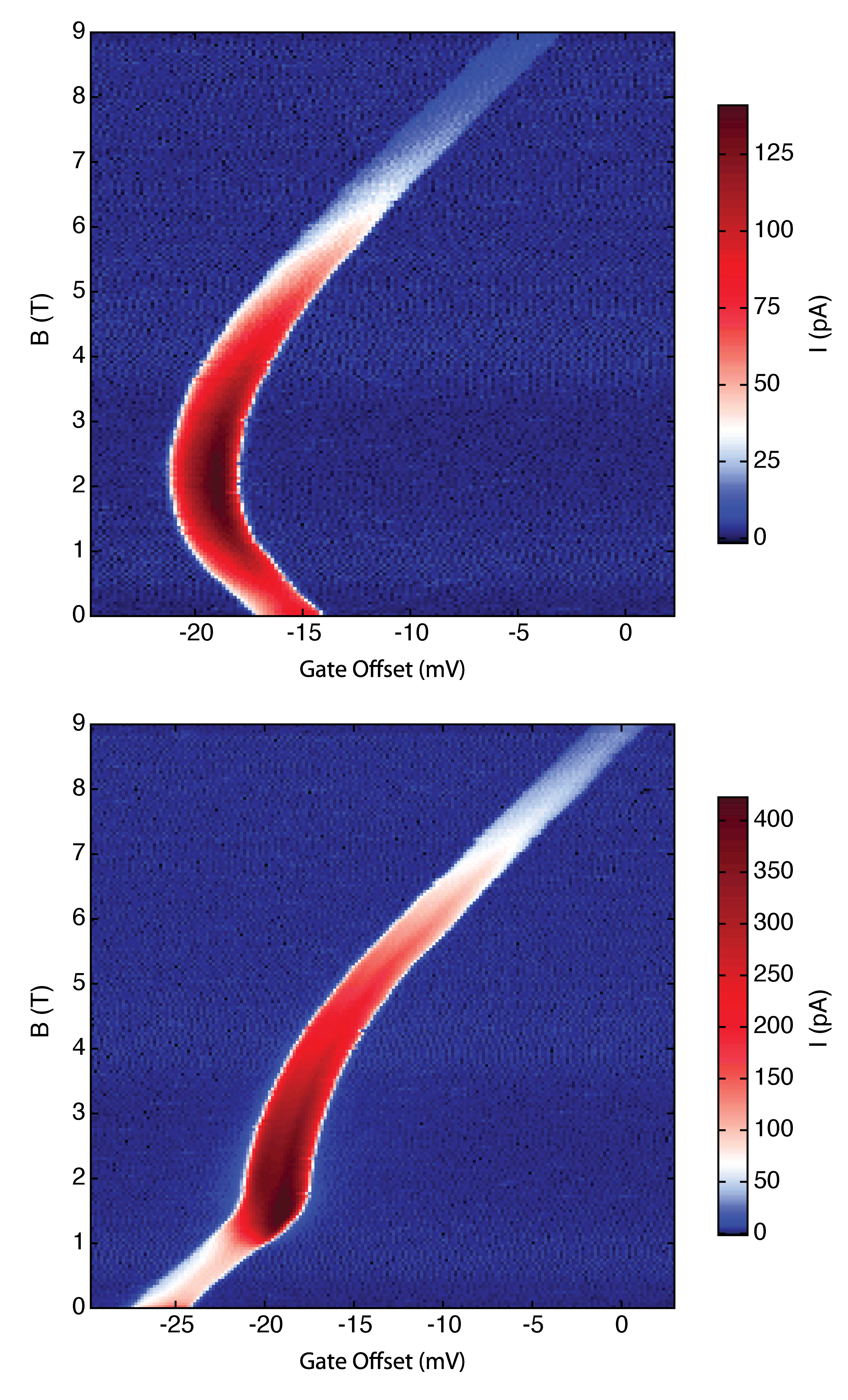}
\end{center}
{\bf Supplementary Figure S10: High resolution datasets in Device 1 of
  spin-orbit states.} High resolution datasets of the 1e and 2e Coulomb
peaks of device 1 with $V_{SD} = 1$ mV, showing behaviour at low
magnetic fields. Here, the low field behaviour is not affected by
artifacts in the first gate sweep first gate sweep at because of a
slow sweep rate of the gate that provided sufficient time for the
magnetic field controller to settle before reaching the position of
the Coulomb peak.

\vspace*{\fill} \newpage \vspace*{\fill}

\begin{center}
\includegraphics[width=0.6\textwidth]{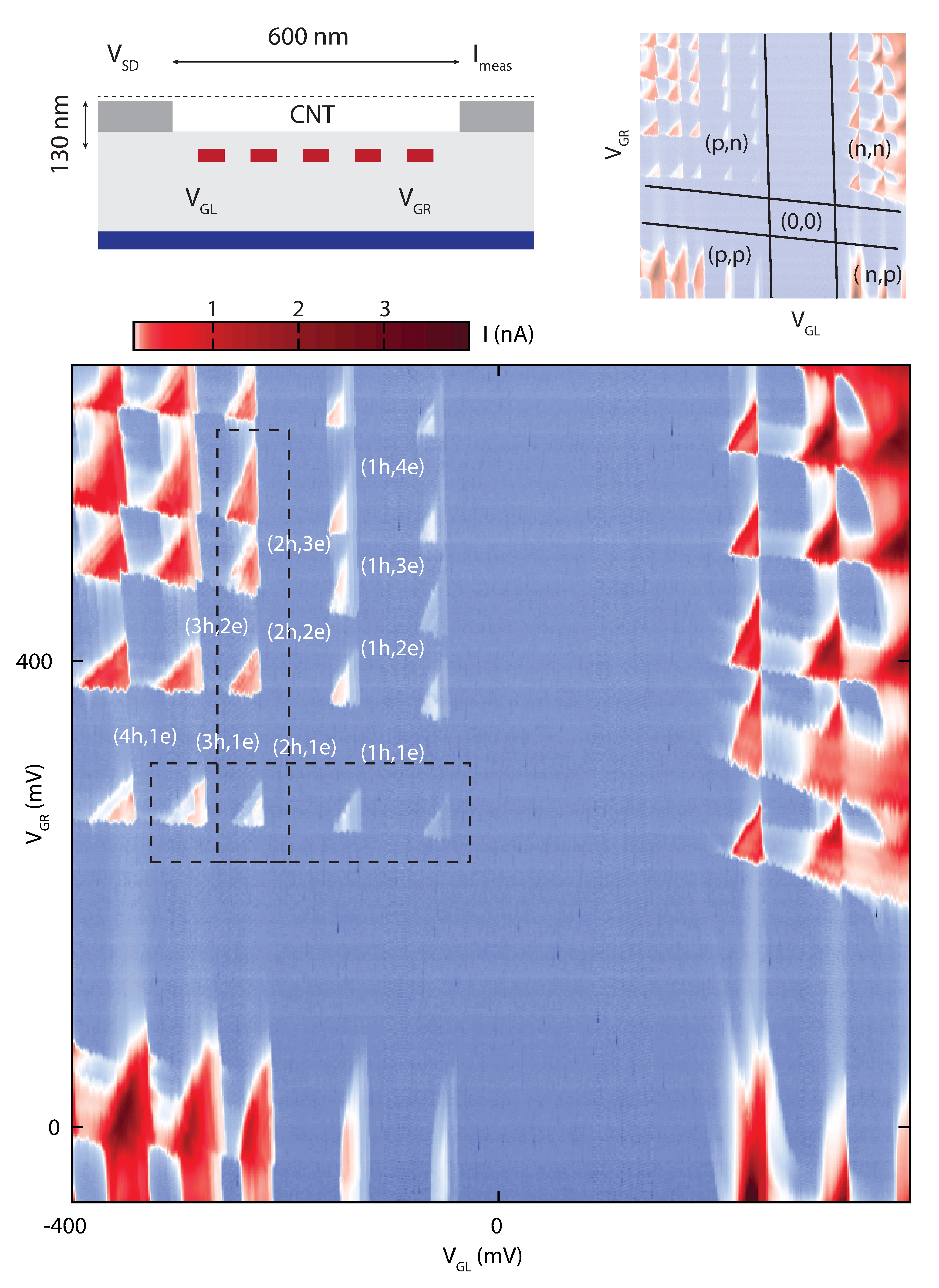}
\end{center}

{\bf Supplementary Figure S11: Characterization of Device 3.} {\bf a,}
Schematic of device 3. The total device length is 600 nm. The device
includes 5 local gates embedded in oxide under the suspended
nanotube. In the measurements, the outermost gates ($V_{GL}$,
$V_{GR}$) are used to tune the electron/hole number in a (p,n) type
double quantum dot, while the inner three gates are used to tune the
interdot tunnel barrier. {\bf b,} An overview of gate space,
indicating the (p,p), (p,n), (n,p), and (n,n) regions of gate space,
and the identification of the (0,0) configuration.  {\bf c,} A color
scale plot of the measured current as a function of $V_{GL}$ and
$V_{GR}$ at $V_{SD} = 10$ mV. The boxes outlined by dashed lines show
the triple points used to track the ground state energies in figure
S12 and figure 4 of the main text.

\vspace*{\fill} \newpage 

\begin{center}
\includegraphics[width=0.8\textwidth]{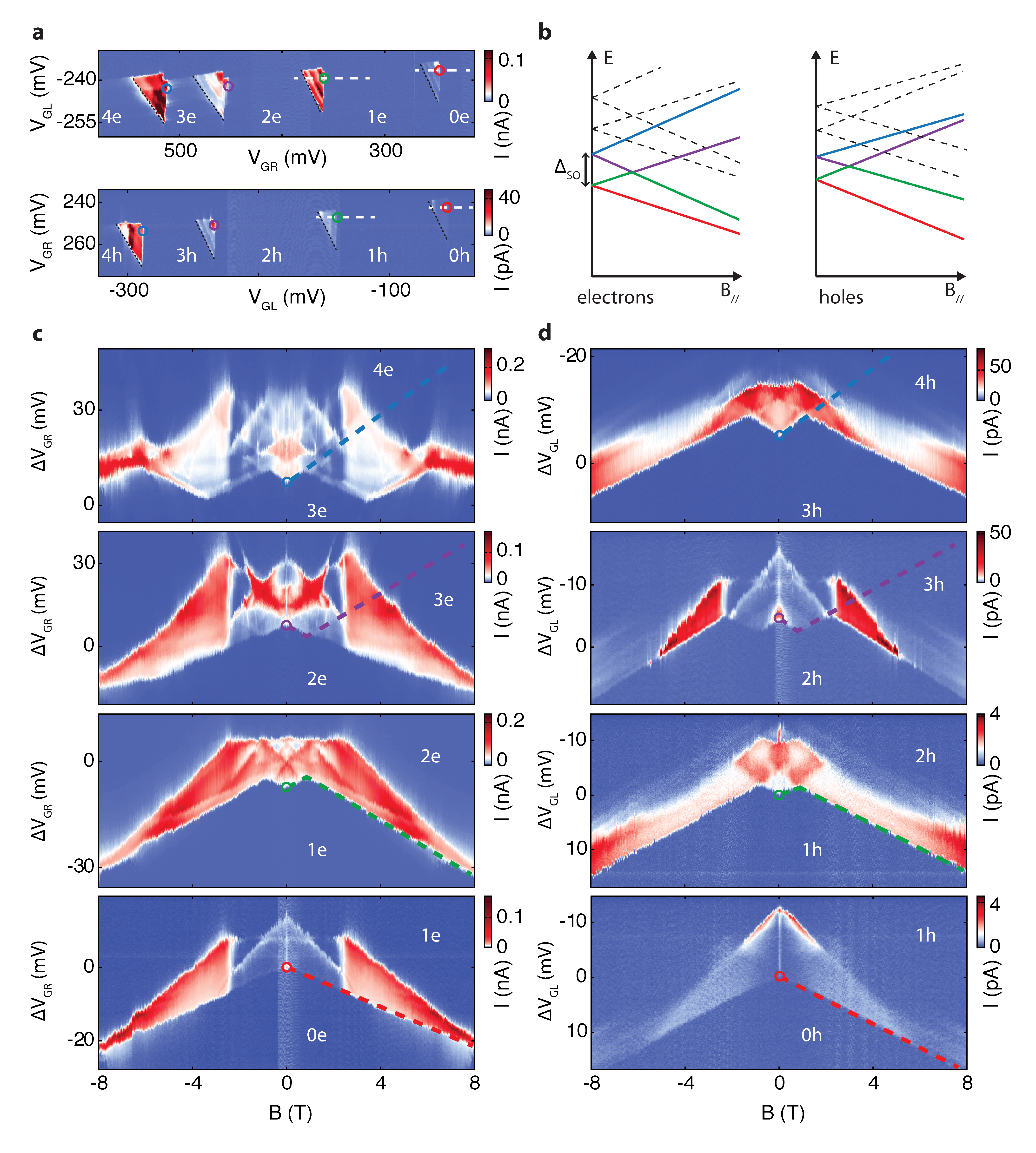}
\end{center}
{\bf Supplementary Figure S12: Energy spectra of the first electron
  shell and hole shell of Device 3.} Extended datasets from figure 4 of
the main text, showing the ground states of the first four electrons
and the first four holes, extracted from the motion of the triple
points indicated in the dashed boxes in figure S11 with magnetic
field. {\bf a,} Double-dot stability diagrams taken at $V_{sd} = 5$
mV. {\bf b,} Expected spectra for the first four electrons (solid
lines), as well as spectra from the next higher shell (dashed
lines). {\bf c,d,} Magnetic field dependence of gate space cuts (white
dashed lines in {\bf a}) for the first four electrons {\bf c} and
holes {\bf d}.  Similar to previous works [8],
the magnetic field dependence of the third and fourth electrons/holes
does not follow exactly the single-shell spin orbit spectrum (solid
lines in {\bf b}), but instead show extra crossings from downward
moving levels in higher shells (dashed lines in {\bf b}).

\newpage \vspace*{\fill}

\begin{center}
\vspace{-1em}
{\bf Supplementary Table S1: Summary of previous spin-orbit
  measurements.}
\vspace{1em}

\begin{tabular}
{ | l | l | l | l | l |}
\hline
Reference & 
$\mu_{orb}$ & 
$d^\star$ &
$\Delta_{SO}^{max}$ theory & $
\Delta_{SO}$ observed 

\\ \hline \hline
Kuemmeth {\em et al.} [8]&
1.55 meV/T & 
7.0 nm & 
110 $\mu$eV &
\parbox[t]{3cm}{
370 $\mu$eV (1e) \\
210 $\mu$eV (1h) \vspace{0.5em}} 

\\ \hline
Jespersen {\em et al.} [8] &
0.63 meV/T & 
2.9 nm & 
168 $\mu$eV &
150 $\mu$eV (many electrons) 

\\ \hline
Jespersen {\em et al.} [31] &
0.87 meV/T & 
5.3 nm$^\dagger$  & 
146 $\mu$eV &
200 $\mu$eV (many electrons) 

\\ \hline
Churchill {\em et al.} [9] &
0.33 meV/T & 
1.5 nm  & 
520 $\mu$eV &
170 $\mu$eV (1e)

\\ \hline
Jhang {\em et al.} [10] &
0.33 meV/T$^{\diamond}$ & 
1.5 nm$^{\star\star}$ & 
520 $\mu$eV &
2500 $\mu$eV$^{\dagger\dagger}$

\\ \hline
Device 1 &
1.6  meV/T & 
7.2 nm  & 
106 $\mu$eV &
3400 $\mu$eV (1e)

\\ \hline
Device 1 &
1.6  meV/T & 
3 nm$^\ddagger$  & 
260 $\mu$eV &
3400 $\mu$eV (1e)

\\ \hline
Device 2 &
1.5  meV/T & 
6.8 nm  & 
116 $\mu$eV &
1500 $\mu$eV (1e)

\\ \hline
Device 3 &
0.9  meV/T & 
4.1 nm  & 
190 $\mu$eV &
1700 $\mu$eV (1e)

\\ \hline
Device 3 &
0.8  meV/T & 
3.7 nm  & 
208 $\mu$eV &
1300 $\mu$eV (1h)

\\ \hline
\end{tabular}

\vspace{1em}
\begin{spacing}{1.5}
\parbox{0.9\textwidth}
{ \small
$^\star$ Estimated from the observed orbital magnetic moment, ignoring
effects of $k_{||}$, unless otherwise noted\\ 
$^\dagger$ The value of
the diameter for this entry is based on a detailed analysis of
$\mu_{orb}$ as a function of shell number performed by the authors. \\
$^\ddagger$ The diameter for this entry is based on the observed AFM
height of the nanotube.\\
$^{\diamond}$ Orbital moment implied from AFM diameter. \\
$^{\star\star}$ Diameter from AFM. \\
$^{\dagger\dagger}$ Implied from bulk bandgap measurements.\\
}
\end{spacing}
\end{center}

\vspace*{\fill} \newpage  

\sn{Supplementary Note 1: Characterization of Device 1}

A schematic of Device 1 is shown in figure S1{\bf a}. Similar to
previous studies[22], we make a clean
suspended carbon nanotube quantum dot by growing the nanotube across a
pre-defined structure in the last step of the fabrication. A SEM image
of the actual device (taken after all measurements were completed) is
shown in figure S1{\bf b}. As we do not control the direction of the
nanotube growth, it often crosses the trench at an angle, as can be
seen in this device. From AFM measurements, we estimate the nanotube
diameter to be 3 nm.

\ind

We apply a d.c. voltage across the source and drain of the device and
measure the current through the nanotube as we sweep the gates, as
shown in figure S1{\bf c}. In the upper left corner of the plot, the
gates dope the center of the nanotube with holes. Near the edge of the
device, the gate electric fields are screened by the ohmic contact
metal; here, the doping is set by the work function difference between
the metal ($\Phi_{Pt} \sim 5.6$ eV) and the nanotube ($\Phi_{CNT} \sim
4.9$ eV), resulting in a gate-independent hole doping at the edge of
the trench. This, combined with hole doping of the suspended segment
from the gates, results in a p$'$pp$'$ configuration in the upper left
corner of figure S1{\bf c}. In this region, we observe only weak
modulations of the conductance which does not vanish between peaks,
indicating a highly transparent interface between the Pt metal and
clean nanotube. In the lower right corner of figure S1{\bf c}, the
gates induce electrons in the suspended segment, giving a p$'$np$'$
doping profile. Electrons occupy a quantum dot with tunnel barriers
defined by p-n junctions [22], in which we can
count the number of carriers starting from zero, shown in figure
S1{\bf e}.

Figure S1{\bf d} shows $I_{meas}$ vs.\ $V_g$ taken along the dotted
line in S1{\bf c} as a function of an external magnetic field applied
in the plane of the sample, perpendicular to the trench.  The distance
in gate voltage between the onset of electron and hole current is a
measure of the electronic bandgap of the nanotube. In carbon
nanotubes, a magnetic field component parallel to the nanotube axis
shifts the quantization condition of the states circling the
circumference ($k_\perp$) by an Arahonov-Bohm flux, and therefore
reduces the nanotube bandgap (see figure 2{\bf a} of main text). For
sufficiently large magnetic fields, the $k_\perp$ quantization line
will cross the Dirac point of the graphene bandstructure and the
bandgap begins to increase again. In our device, this occurs at a
magnetic field of $B_{Dirac} = 2.2$ T, indicated by white arrows in
figure S1{\bf d}. This implies a contribution to the electronic
bandgap $E_{gap}^{k\perp} = 2 \hbar v_F \Delta k_\perp \sim 7$ meV
arising from the shift of the $k_\perp$ quantization line. In this
sense, our nanotube is very close to the metallic condition in which
the $k_\perp$ quantization line passes directly through the center of
the Dirac cone. This is a very different regime compared to previous
devices where the nanotube spin orbit coupling was studied [8,11], in
which no such evidence of a low Dirac field was seen. Similar to
previous studies where low Dirac fields were reported [28], the
bandgap do not vanish at the Dirac point. We observe a residual gap in
the transport data at the Dirac point of about 80 meV, measured by
subtracting the average of the addition energies from the first
electron and the first hole from the addition energy of the empty
quantum dot.

\newpage
\sn{Supplementary Note 2: Spin orbit splitting in Device 2}

In figures S6-S9, we present the magnetic field dependence of the
ground states of the first four electrons in a second nearly metallic
carbon nanotube (device 2). Device 2 is similar in design to device 1,
but includes only a backgate. The trench length is 800 nm. In device
2, we observe a Dirac field of 0.8 T, an orbital magnetic moment
$\mu_{orb} = 1.5$ meV/T, and a spin orbit splitting $\Delta_{SO} =
1.5$ meV.

\newpage
\sn{Supplementary Note 3: Model for a nearly metallic nanotube with spin-orbit coupling}

In order to calculate the spectra plotted in figures 2{\bf g} and {\bf h} of
the main text, we use a model of the nanotube based on the
graphene bandstructure with a parallel magnetic field. In a basis of
spin and valley eigenstates in which the spin direction is defined
parallel to the axis of the nanotube, the Hamiltonian consists of a
4x4 matrix with only diagonal elements given by:
\begin{equation}
E(v,s,B) = \sqrt{(E_{k\perp}+vs\Delta_{SO}^{orb}+v\mu_{orb}B)^2+
   E_{k\parallel}^2} + vs\Delta_{SO}^{Zeeman} + \frac{1}{2}sg\mu_BB
\end{equation}
Here, $v$ and $s$ take on values $\pm 1$ depending on the electron
spin and the valley it occupies, $E_{k(\parallel,\perp)} = \hbar v_F
k_{(\parallel,\perp)}$ where $k_{(\parallel,\perp)}$ are the momentum
of the electron relative to the Dirac points in the directions
parallel and perpendicular to the axis of the nanotube, and
$\Delta_{SO}^{orb}$ and $\Delta_{SO}^{Zeeman}$ are the orbital and
Zeeman type spin orbit splittings at $k_{||} = 0$ ($\alpha$ and
$\beta$). These diagonal elements correspond to the energies plotted
in figure 3. In the calculations, we have chosen to make the total spin
orbit coupling either purely orbital or purely Zeeman for illustrative
purposes, and have used the following parameters: $\Delta_{SO} = 2 $
meV, $E_{k\parallel} = 1$ meV, $E_{k\perp} = 2$ meV, and $\mu_{orb} =
0.9$ meV/T.

\ind

Including the observed 48 degree misalignment of the magnetic field to
the nanotube axis, the Zeeman splitting Hamiltonian
$g\mu_B \vec{B} \cdot \vec{S}$ is no longer diagonal in this basis, and
the eigenstates are mixtures of the four basis states described
above. However, because the Bohr magneton is small compared to the
orbital magnetic moment, this effect is weak and does not result in
qualitative different spectra.

The Zeeman-type contribution to the spin-orbit splitting, according to
current theoretical estimates, is expected to be larger than the
orbital-type contribution by as much as a factor of 4, except for in
nanotube chiralities where it vanishes or is small due to the
$\cos(3\theta)$ term ($\theta = 0$ corresponding to a zigzag
nanotube). It is an open question, however, why the spin-orbit
splitting we observe in devices 1 and 2 is so dominantly of the
Zeeman-type, with little indication of an orbital contribution.

\newpage
\sn{Supplementary Note 4: Discussion of summary table of previous spin-orbit spliting measurements}

In Supplementary Table S1, we summarize in a table our measurements
together with other measurements of the spin-orbit coupling reported
in literature. As described in the main text, we use the formula for
the nanotube spin-orbit splitting from [27], given by:
\begin{equation}
H^{cv}_{SO} = \alpha S^z \sigma_1 + \tau \beta S^z
\end{equation}
with the orbital contribution given by:
\begin{equation} 
\Delta_{SO}^{orb} = \alpha = \frac{-0.08\ \textrm{meV\ nm}}{r}
\end{equation}
and the Zeeman contribution given by:
\begin{equation} 
\Delta_{SO}^{Zeeman} = \beta = \frac{-0.31\ \textrm{meV\ nm}}{r}
\end{equation}
where $r$ is the radius of the nanotube. For the maximum theoretical
value, we choose $\theta = 0$, giving:
\begin{equation}
\Delta_{SO}^{max} = \frac{780\ \mu\textrm{eV}}{d\ (\textrm{in nm})} 
\end{equation}
 
\ind 

In order to provide a consistent comparison, we have estimated the
(minimum) diameter using the observed value of the orbital magnetic
moment $\mu_{orb}$. Assuming a Fermi velocity of $0.9 \times 10^{6}$
m/s, $\mu_{orb}$ is given by:
\begin{equation}
\mu_{orb} = \frac{d e v_F}{4} = 220\ \mu\textrm{eV / T} \times d\ (\textrm{in nm})
\end{equation}
where $v_F$ is the Fermi velocity of the graphene bandstructure, which
we take here as $0.9 \times 10^6$ m/s. Here, we assume $v_{F\perp} =
v_F$, and therefore have not accounted for the reduction of $\mu_{orb}$
from a finite $k_{||}$ [31]. The resulting estimates
of $d$ from $\mu_{orb}$ represent a lower bound on the diameter (and
thus also an upper bound on $\Delta_{SO}^{max}$).

We have also included three entries in which we calculate
$\Delta_{SO}^{max}$ based on a different estimate of the
diameter. These three entries correspond to the diameter $d = 3$ nm we
estimate from AFM measurements on device 1, the diameter $d = 5.3$ nm
estimated by [31] from an extensive analysis of
$\mu_{orb}$ as a function of gate voltage, and the diameter $d = 1.5
$ nm measured by Jhang {\em et al.} [10]. Note that the
tapping-mode AFM measurement of the diameter may underestimate the
diameter of single-wall carbon nanotubes due to compression forces
from the AFM tip.

The measurements referred to in the table were performed by tracking
the electronic states of individual levels in a quantum dot at low
temperatures, except for the measurements of Jhang {\em et
  al.}\ [10]. The values in [10] are based on measurements of the
nanotube bandgap implied from device conductance near the bandgap as a
function of magnetic field at different fixed gate voltages, together
with the nanotube diameter as measured by AFM.
The devices measured here and those measured by Kuemmeth {\em et al.}\ [8]
were made using clean nanotubes grown in the last
step of the fabrication, while the other measurements were performed
on nanotubes which were grown first and subsequently underwent
processing in the cleanroom.

Finally, we also note that when using $\mu_{orb}$ to estimate the
nanotube diameter, we obtain a number that is not only larger than the
AFM measurement for device 1, but also larger than the largest
diameter expected for single wall carbon nanotubes in, for example,
transmission electron microscope studies. This is also the case for
many of the devices in Table 1. Such a discrepancy was also noted by
earlier authors [31], and remains unresolved. One suggestion of the
authors of [31] was a renormalization of the Fermi velocity. Such a
renormalization could arise from, for example, discrepancies between
the experimental tight binding parameters of carbon nanotubes and
those obtained from {\em ab initio} calculations.

\newpage
\sn{Supplementary Note 5: Artifacts in extracted ground state energies at $B < 0.15$ T}

Note that there is glitch in the first line of the data set in figure
S6.  This artifact is also present to a lesser degree figures 1(c)-(f)
and the resulting extracted energies in figures 3(a)-(c) of the main text.
This glitch results in an artifact in the resulting extracted ground
state energies plotted in figure S5 in the form of a flat slope for
$B<150$ mT.  The glitch and resulting artifacts arise from the
inability of our magnetic controller to track the setpoint field
during faster magnetic field sweeps. The effects of these artifacts
are limited to the first gates sweep (row) of the Coulomb peak
magnetic field dependence data. These artifacts have been accounted
for in the estimation of the error bar on $\Delta_{SO}$.

\ind

In order to demonstrate that these artifacts are not obscuring
possible other phenomena at very low magnetic fields, we have also
included high resolution datasets in figure S10 for the data in figure
1(c) and 1(d) of the main text. Here, the gate was swept sufficiently
slowly that the magnet controller had time to settle before the gate
voltage reached the position of the first Coulomb peak, and thus the
artifacts are not present.

\newpage

\sn{Supplmentary Note 6: Device 3 characterization and analysis}

In this section, we present a basic characterization of device 3
(figure S11), together with measurements the magnetic field dependence
of the ground state energies of the first four electrons and first
four holes in the device (figure S12), and discuss the extraction of
the ground state energies from the magnetic field dependence of the
gate-space cuts through the triple-point triangles.

\ind 

By tracking the gate voltage position of any fixed point on the
triple-point bias triangles as a function of magnetic field, we can
independently track the ground state energy of the left and right dot
in the double quantum dot device. This is analogous to the tracking of
the ground states of a single quantum dot by following the Coulomb
peak position with magnetic field. To make this concrete, we
illustrate this in the context of upper left bias triangle in figure
4{\bf a} of the main text, corresponding to the (3h,1e)
$\leftrightarrow$ (2h,0e) transition. In the case that there is very
small crosstalk capacitance from the left gate to the right dot (as is
the case in figure 4{\bf a} of the main text where the edges of the
triple-point bias triangle are nearly vertical), vertical shifts of
the bias triangle arise from shifts in the 3h ground state, while
shifts in the 1e ground state shift the bias triangle horizontally. In
measuring the shift of the bias triangle, it is equivalent to track
any fixed point on the triangle. We choose to extract the ground state
energies by following a point near the tip of the triangle, as the
current on the baseline in our device is weak due to weakly
tunnel-coupled ground states.

\newpage 

\sn{Supplementary References}
\vspace{-1em}

\begin{enumerate}
\expandafter\ifx\csname url\endcsname\relax
  \def\url#1{\texttt{#1}}\fi
\expandafter\ifx\csname urlprefix\endcsname\relax\def\urlprefix{URL }\fi
\providecommand{\bibinfo}[2]{#2}
\providecommand{\eprint}[2][]{\url{#2}}

\setcounter{enumi}{33}
\renewcommand{\labelenumi}{[\arabic{enumi}]}

\item
\bibinfo{author}{Steele, G.} \emph{et~al.}
\newblock \bibinfo{title}{Strong coupling between single-electron tunneling and
  nanomechanical motion}.
\newblock \emph{\bibinfo{journal}{Science}} \textbf{\bibinfo{volume}{325}},
  \bibinfo{pages}{1103} (\bibinfo{year}{2009}).

\end{enumerate}
